\documentclass{article}

\usepackage[english]{babel}

\usepackage[letterpaper,top=2cm,bottom=2cm,left=3cm,right=3cm,marginparwidth=1.75cm]{geometry}

\usepackage{amsmath}
\usepackage{graphicx}
\usepackage[colorlinks=true, allcolors=blue]{hyperref}
\usepackage{booktabs}
\usepackage{float}
\usepackage{longtable}
\usepackage{tabularx}
\usepackage{caption}
\usepackage{multirow}  
\usepackage{authblk}
\usepackage{amsmath}
\usepackage{subcaption}
\usepackage{cite}

\title{The Geography of Transportation Cybersecurity: Visitor Flows, Industry Clusters, and Spatial Dynamics}

\author[1]{\small Yuhao Wang}
\author[1]{Kailai Wang\thanks{Corresponding author: kwang43@central.uh.edu}}
\author[2]{Songhua Hu}
\author[3]{Yunpeng(Jack) Zhang}
\author[1]{Gino Lim}
\author[4]{Pengyu Zhu}

\affil[1]{Department of Industrial and Systems Engineering, University of Houston, Houston, TX 77024, USA}
\affil[2]{Senseable City Lab, Department of Urban Studies and Planning, Massachusetts Institute of Technology, Cambridge, MA, 02139, USA}
\affil[3]{Department of Information Science Technology, University of Houston, Houston, TX 77024, USA}
\affil[4]{Division of Public Policy, Hong Kong University of Science and Technology, Hong Kong 511455, CN}
\begin{document}
\maketitle

\noindent\textbf{Abstract:} The rapid evolution of the transportation cybersecurity ecosystem—encompassing cybersecurity, automotive, and transportation and logistics sectors—will lead to the formation of distinct spatial clusters and visitor flow patterns across the US. This study examines the spatiotemporal dynamics of visitor flows, analyzing how socioeconomic factors shape industry clustering and workforce distribution within these evolving sectors. To model and predict visitor flow patterns, we develop a BiTransGCN framework, integrating an attention-based Transformer architecture with a Graph Convolutional Network (GCN) backbone. Using SafeGraph’s one-year weekly visitor flow data along with additional socioeconomic indicators, we conduct a multi-scale spatial clustering analysis to identify distinct industry concentration trends, workforce mobility patterns, and geographic interdependencies across three key transportation cybersecurity-related industries. Our findings reveal that geolocation and education levels are the most significant factors influencing industry cluster formation. Furthermore, our predictive model anticipates a 14.16\% average increase in visitor flow across the US in the next term, highlighting emerging industry growth, regional workforce shifts, and cross-sector innovation. By integrating AI-enabled forecasting techniques with spatial analysis, this study improves our ability to track, interpret, and anticipate changes in industry clustering and mobility trends, thereby supporting strategic planning for a secure and resilient transportation network. It offers a data-driven foundation for economic planning, workforce development, and targeted investments in the transportation cybersecurity ecosystem.
\vspace{2\baselineskip}

\noindent\textbf{Keywords:} Transportation Cybersecurity, Visitor Flows, Spatial Clustering, BiTransGCN model, Social Impacts

\section{Introduction}

The increasing integration of connected and automated vehicles (CAVs) into the automotive industry has heightened its reliance on technology, exposing vehicles to growing cybersecurity risks, including hacking, data breaches, and system malfunctions \cite{khan2023cybersecurity, thomas2025cybersecurity}. However, cybersecurity challenges in transportation are not isolated threats; they should be understood within a broader, interdependent ecosystem spanning the cybersecurity, automotive, and transportation and logistics industries. Addressing these challenges requires a holistic, spatially aware analytical approach that accounts for regional variations and industry linkages. Thus, this study analyzes spatiotemporal dynamics of interconnected industries across U.S. regions using origin-destination (OD) visitor flow data to identify interdependencies, assess regional vulnerabilities, and inform cybersecurity strategies.
The benefits of geographically concentrated industrial clusters are adequately documented. Such clustering reduces business costs and enhances efficiency \cite{rosenfeld1997bringing}. Analyzing the underlying sources of cluster externalities provides valuable insights into the key patterns of economic development shaped by industrial agglomeration.

Researchers have explored industrial clusters from various perspectives, such as the role of externalities in advancing high-tech industries \cite{choi2019high}, the development of regional clusters for innovation \cite{fundeanu2014impact}, and global value chain clusters from an economic geography perspective \cite{boschma2022global}. Yet, one critical perspective remains underexplored: the impact of cross-regional labor mobility and visitor flow in shaping industrial clusters. Our research seeks to fill this gap by investigating the evolution of transportation cybersecurity industry (TCI) clusters. 
Business visitor flow is essential economically, promoting knowledge spillovers and improving workforce management. It enhances regional collaboration and economic growth, fostering social networks across businesses (e.g., \cite{neffke2017inter}). On a deeper level, the visitor flow-based TCI is closely tied to broader economic environments, social conditions, education, healthcare, and other societal factors, all of which exhibit variations across time and space \cite{jones2012social,bejjani2023digital,elekes2024regional}. This study employs spatiotemporal models and deep learning techniques to examine how TCI-related industries cluster over time and space. We attempt to answer the following questions: (1) How the spatiotemporal mobility of TCI visitor flows in the US is influenced by social factors, including regional economies, security, education, health, housing, and work conditions? (2) What are the spatiotemporal characteristics of clustering among the three TCI industries: cybersecurity, automotive, and transportation and logistics? (3) How can the deep learning model effectively predict the proportional changes of visitor flow across geographical regions? By answering these questions, this study contributes to literature in two aspects. First, to the authors’ knowledge, it is the first to focus on TCI clusters from a visitor flow perspective and reveal the underlying connections between cluster levels and various social and economic factors. Second, we design a hybrid deep learning-based approach that combines Transformer and GCN to predict spatiotemporal variations in the long-term visitor flow. The findings offer actionable insights into industrial cluster dynamics, guiding policymakers in resource allocation and strategic planning.

\section{Literature Review}

\subsection{Industry Spatiotemporal Clustering Pattern}

Clusters—geographic concentrations of related industries and supporting institutions—represent a fundamental concept in economic geography. Marshall (1920) identified three main drivers: input-output linkages, shared labor markets, and knowledge spillovers, all of which contribute to cost savings and increased productivity \cite{marshall1920principles}. Subsequent research has broadened this framework to include additional factors such as localized demand conditions, specialized institutions, regional business structures, and embedded social networks \cite{saxenian2006new,markusen2017sticky,sorenson2000social}. Essentially, clusters foster industry connections through shared knowledge, skills, market demand, and institutional support, promoting growth and collaboration \cite{weisbrod2014evolving}. Clusters emerge from the interplay of first-mover advantages, economies of agglomeration, localized technology sharing, and geographic path dependencies, forming the foundation of cluster-based economic strategies. Researchers have identified key drivers of agglomeration economies, including input-output relationships, shared labor markets, knowledge diffusion, local demand, specialized institutions, and business and social networks \cite{marshall1920principles,porter1990comparative}. Beyond economic factors, social influences also shape industrial clustering. Florida (2002) emphasized the role of "place quality" in attracting and retaining talent, while access to transportation is critical for supporting technology-driven businesses \cite{florida2002economic}. Firms relocating away from clusters risk losing the competitive advantages of a concentrated network, including shared resources, infrastructure, and collaboration.

Economic geography explains that industrial and economic activities are unevenly distributed across time and space, with clustering occurring in specific locations or periods \cite{elekes2024regional}. The timing and location of economic activities are interconnected, contributing to regional disparities in industrial development \cite{massey1995spatial}. In addition to cost advantages, clustering is driven by "non-traded interdependencies" that sustain the competitive edge of cities and regions \cite{storper1997regional}. Factors such as regional economic structures, education levels, social security systems, and healthcare infrastructure collectively shape spatiotemporal clustering dynamics. A major research focus is identifying the effects of industrial clusters and understanding their geographic distribution patterns. Some scholars examine input-output relationships and value chain connections, while others analyze geocoded data to uncover hidden cluster dynamics \cite{rosenfeld2022spatial}. Debates persist in assessing cluster size, critical mass, and growth variations. Leslie and Kronenfeld (2011) used the location quotient to measure specialization \cite{leslie2011colocation}, while Liu et al. (2021) \cite{liu2021detecting} proposed co-location pattern mining to identify multi-industry relationships. Recent studies integrate spatiotemporal models with global economic data to quantify spatial-temporal interactions \cite{kim2021spatial}. While some research prioritizes firm interdependence without geographic considerations, others emphasize geographic concentration as a crucial factor in understanding industrial clusters, providing valuable economic policy and strategy insights.

Research on TCI clustering processes can significantly contribute to understanding and addressing key challenges in the realms of Next-Generation Transportation Cybersecurity Systems \cite{khan2023cybersecurity,thomas2025cybersecurity}. These clusters often involve diverse stakeholders with expertise in various domains such as data analytics, artificial intelligence, cryptography, and network security \cite{paramesha2024big}. Although previous studies have acknowledged the significance of industrial clustering and spatial pattern characteristics, they have not explored the cluster effects of TCI from the perspective of visitor flow. There is also a lack of quantitative analysis on the correlation between social and economic factors and clustering levels. Additionally, existing research on transportation cybersecurity primarily focuses on attack and defense mechanisms for transportation data, with limited attention to the long-term prediction of TCI’s spatial distribution dynamics.

\subsection{Deep Learning-Based Long-Term Flow Prediction}
Long-term traffic forecasting is essential for gaining deeper insights into how talent moves across regions and industries over time, which is a valuable area to explore. However, work within the scope of long-term traffic flow prediction is relatively rare. Traffic prediction remains a challenging task due to the dynamic nature of traffic conditions and the complex spatiotemporal correlations inherent in traffic data. Traditional forecasting methods—such as the Historical Average (HA) model, Vector Auto-Regression (VAR) \cite{zivot2006vector}, and Auto-Regressive Integrated Moving Average (ARIMA) \cite{zhang2003time}-have been widely applied in traffic forecasting. While these models offer useful baseline predictions, they often struggle to capture the nonlinear and dynamic patterns present in real-world traffic systems.

With the rapid advancement of deep learning technology, various models have been developed for traffic prediction. Convolutional Neural Networks (CNNs) identify spatial patterns in grid-based traffic networks, while Recurrent Neural Networks (RNNs) capture temporal dependencies \cite{wu2016short}. In contrast, many researchers have advocated for using domain-specific attention mechanisms, with Transformers emerging as a powerful architecture for traffic flow prediction \cite{yan2021learning}. The models rely on an encoder-decoder structure, where each encoder layer focuses on capturing the relationships among different elements of input data. The self-attention mechanism allows these models to flexibly capture long-range dependencies and effectively learn sequential patterns and improve their ability to analyze complex relationships \cite{yan2021learning,reza2022multi}. CNNs and other deep learning models have been widely used to model spatial dependencies in data \cite{zhang2016dnn}. However, these models are limited to Euclidean data structures and cannot adequately address the complexities of non-Euclidean road networks. In contrast, GNNs offer a more robust approach for cross-regional spatiotemporal modeling, effectively handling irregular data structures and uncovering latent features for long-term geographic mobility predictions \cite{li2023augmentation,wang2024identification}. In traffic flow prediction, researchers have leveraged static traffic network topologies as spatial graphs within GNN architectures, enabling message-passing mechanisms to propagate traffic information. Models such as Spatiotemporal GCN \cite{yu2017spatio} and Temporal GCN (T-GCN) \cite{zhao2019t} have demonstrated strong predictive capabilities, further refined by approaches such as spatial–temporal synchronous GCN \cite{song2020spatial} and structure learning convolution (SLC) \cite{zhang2020spatio}. Recent innovations include STGC-GNNS \cite{he2023stgc}, which enhances long-term forecasting accuracy, and spatiotemporal dual adaptive GCN (ST-DAGCN) \cite{liu2024st}, which uses adaptive adjacency matrices to capture both global trends and local dynamics. 

The combination of a Transformer based on attention mechanisms with GCN can simultaneously capture both the local topological structure and global dependencies in the graph data, enhancing the model's understanding of complex features and potential associations in the data \cite{wang2023sat}. The self-attention mechanism helps the model flexibly establish long-range relationships between feature nodes, acquiring global feature information \cite{ashish2017attention}. Meanwhile, GCN efficiently aggregates local neighborhood information, providing a multi-level approach to information fusion \cite{kipf2016semi}. The dual-model approach has stronger generalization capabilities, making it suitable for long-term traffic flow prediction tasks, with more accurate expression of node associations and predictions of spatiotemporal variations.

Overall, the literature review reveals several significant research gaps. The TCI should be understood within a broader and interdependent ecosystem. However, existing studies lack a comprehensive and spatially aware framework for analyzing its spatiotemporal clustering. Additionally, there is a need for a more advanced and effective long-term prediction method to assess changes in visitor flows accurately. The impact of social and economic factors on TCI clustering requires deeper investigation to address challenges related to industrial innovation and economic growth in the U.S. With all these in mind, we hypothesize: (1) three industries within TCI (cybersecurity industry, automotive industry, and transportation and logistics industry) exhibit different spatial distributions and clustering trends, with varying sensitivity to socio-economic factors; (2) a hybrid model combining Transformer and GCN can effectively predict long-term changes in visitor flow distribution; and (3) social and economic variables impact TCI clustering levels to different extents, with geographic location and education level emerging as the most influential factors in overall industrial clustering.

\section{Data and Study Area}
\subsection{Data Processing}
This study investigates the spatiotemporal dynamics and socioeconomic aspects of the TCI-related industries in the US using OD-based visitor flow data collected in 2022 by SafeGraph. The dataset provides detailed information for each location, including its name, address, geographic coordinates (latitude/longitude), timestamps, North American Industry Classification System (NAICS) codes, and industry categories. Additionally, we incorporate socioeconomic variables from the American Community Survey (ACS) to contextualize the analysis. SafeGraph’s Point of Interest (POI) dataset leverages extensive smartphone location data to track individual movements, enabling location-based analysis. It records the number of visitors to specific POIs within a given time frame, along with their home census block group (CBG) information. Using this data, we generate information on the number of visitors traveling from specific CBGs to particular POIs within a one-week period. Following data collection, we implement a structured data processing pipeline, which includes preprocessing, classification, statistical analysis, and visual presentation of results. During preprocessing, millions of records are read and converted into a structured dataframe format. Table labels and corresponding data items are carefully verified and aligned to ensure consistency. To maintain data quality and reliability, we manually remove any invalid entries, which we define as empty values or out-of-range data points.

\begin{figure}[H]
    \centering
    \includegraphics[width=1\linewidth]{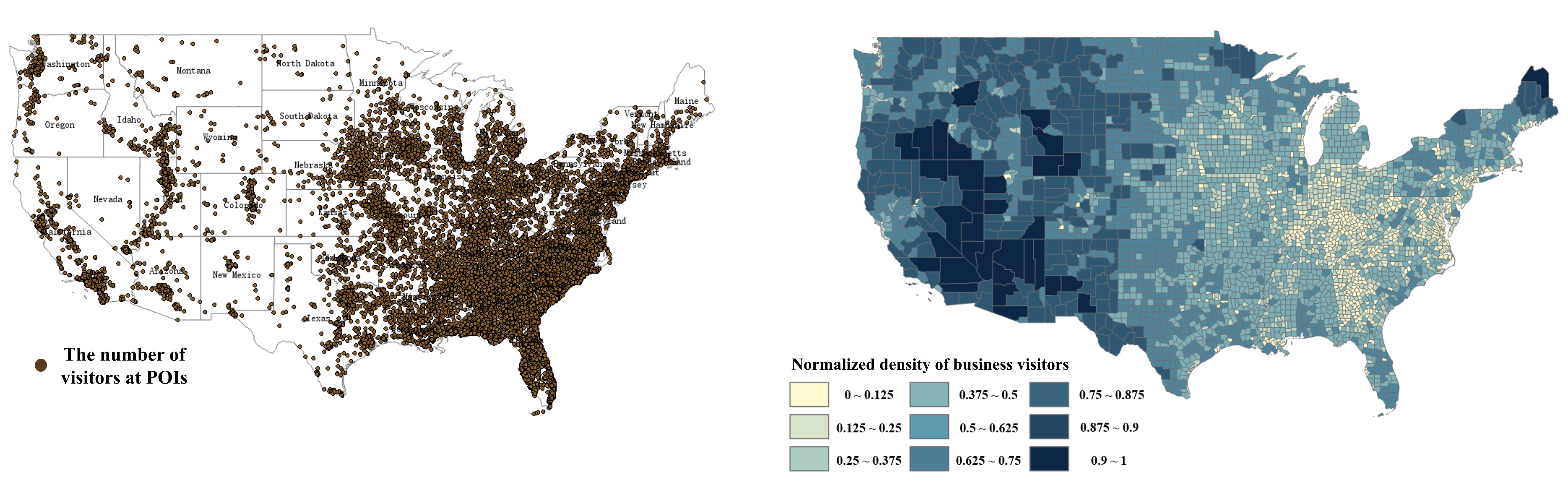}
    \caption{Geographic coding and CGB density maps}
    \label{fig:1}
\end{figure}

We categorize TCI-related visitor flows based on three NAICS code classifications: Automotive Industry Sectors, Cybersecurity Industry Sectors, and Transportation and Logistics Industry Sectors (Table \ref{tab:1}). For each category, we calculate the total number of entities and use this as a representative measure for that industry group. To enhance spatial analysis, we transform the original dataset into digital maps using the geospatial locations of POIs, as illustrated in Fig. \ref{fig:1}. As documented in the literature (Table \ref{tab:2}), socioeconomic factors such as health, education, crime, employment, economic conditions, and housing collectively influence the spatial clustering of industries. Each factor contributes to regional attractiveness, workforce availability, and overall economic viability. Using the ACS datasets, we compute the normalized mean and standard deviation for each variable, providing a comprehensive summary of socioeconomic characteristics in Table \ref{tab:3}.

\begin{table}[H]  
\centering
\caption{\label{tab:1}Visitor data category table in TCI sectors}
\begin{tabular}{ll}
\toprule
\textbf{NAICS code} & \multicolumn{1}{c}{\textbf{Title}} \\
\midrule
\multicolumn{2}{l}{\textit{Automotive Industry Sectors}} \\
336   & Transportation Equipment Manufacturing \\
4231  & Motor Vehicle and Motor Vehicle Parts and Supplies Merchant \\
      & Wholesalers \\
8111  & Automotive Repair and Maintenance \\
\addlinespace
\multicolumn{2}{l}{\textit{Cybersecurity Industry Sectors}} \\
54151  & Computer Systems Design and Related Services \\
541512 & Computer Systems Design Services \\
541519 & Other Computer Related Services \\
56162  & Security Systems Services \\
561622 & Locksmiths \\
541690 & Other Scientific and Technical Consulting Services \\
\addlinespace
\multicolumn{2}{l}{\textit{Transportation and Logistics Industry Sectors}} \\
481 & Air Transportation \\
482 & Rail Transportation \\
483 & Water Transportation \\
484 & Truck Transportation \\
485 & Transit and Ground Passenger Transportation \\
488 & Support Activities for Transportation \\
492 & Couriers and Messengers \\
493 & Warehousing and Storage \\
\bottomrule
\end{tabular}
\end{table}


\begin{table}[htbp]
\centering
\caption{\label{tab:2}Summary of social factors and their impact on industrial clustering}
\begin{tabular}{lll} 
\toprule
\textbf{Social Factor} & \textbf{Impact on Industrial Clustering} & \textbf{Citation} \\
\midrule

Health   & Attracts industries needing healthy workforce. & Oyewole and Thopil (2023)\cite{oyewole2023data} \\
Education & Draw knowledge-intensive industries via skilled labor. & Doeringer and Terkla (1996) \cite{doeringer1996industries}\\
Crime    & Low crime rates enhance safety, attracting businesses. & Fe and Sanfelice (2022) \cite{fe2022bad}\\
Work     & Labor availability drives clustering for talent access. & Chrisinger et al. (2015) \cite{chrisinger2015industry}\\
Economy  & Strong economies attract via low costs and good infrastructure. & Marston (2001) \cite{marston2001effects}\\
Housing  & Affordable housing aids worker attraction, supports clustering. & Kou et al. (2021) \cite{kou2021effect}\\
\bottomrule
\end{tabular}
\end{table}

\begin{longtable}{p{2.6cm} p{1.5cm} p{5cm} p{2.4cm} r r}
\caption{\label{tab:3}Descriptive statistics of social variables sample} \\
\toprule
\textbf{Item} & \textbf{Abbrevia-\allowbreak tion} & \textbf{Description} & \textbf{Code} & \textbf{Mean} & \textbf{Std} \\
\midrule
\endfirsthead

\multicolumn{6}{l}{\textit{(Continued from previous page)}} \\
\toprule
\textbf{Item} & \textbf{Abbrevia-\allowbreak tion} & \textbf{Description} & \textbf{Code} & \textbf{Mean} & \textbf{Std} \\
\midrule
\endhead

\midrule
\multicolumn{6}{r}{\textit{(Continued on next page)}} \\
\endfoot

\bottomrule
\endlastfoot

Poor physical health & PPD & Poor physical health days & v038\_rawvalue & 0.461 & 0.284 \\
Poor mental health & PMD & Average number of poor mental health days & v042\_rawvalue & 0.294 & 0.315 \\
Food environment index & FEI & Index of factors that contribute to a healthy food environment, from 0 (worst) to 10 (best) & V133\_rawvalue & 0.659 & 0.232 \\
High School Education & HSE & \% of Population 25+ with High School (or equivalent) & B15002\_calc\allowbreak\_pctHSE & 0.505 & 0.169 \\
Bachelor’s or Higher Education & BHD & \% of Population 25+ with Bachelor’s or Higher & B15002\_calc\allowbreak\_pctGEBAE & 0.280 & 0.165 \\
Childcare cost burden & CCB & Childcare cost burden (\% of income) & v171\_rawvalue & 0.393 & 0.192 \\
Violent crime offense & VCO & Violent crime offenses per 100,000 & v043\_rawvalue & 0.230 & 0.162 \\
Number of fatalities & FF & Firearm fatalities per 100,000 & v148\_rawvalue & 0.160 & 0.258 \\
Number of injury deaths & NID & Injury deaths with underlying cause & v135\_rawvalue & 0.233 & 0.206 \\
Commute time & CMM & \% that commute more than 30 minutes & v137\_rawvalue & 0.402 & 0.240 \\
Employed population & EP & Employed Population in Civilian Labor Sector & B23025\_004E & 0.866 & 0.168 \\
Working hours & HWP & Mean hours worked (Age 18–64, past 12 months) & B23020\_001E & 0.153 & 0.189 \\
GDP & GDP & GDP from Professional, Scientific, and Technical Services & year\_2019\_60 & 0.110 & 0.170 \\
Median household income & MHI & Median household income (source: income) & V083\_rawvalue & 0.443 & 0.250 \\
Insurance coverage & HIC & \% with Health Insurance Coverage & SUM\_B27010\_calc\allowbreak\_pctNoInsE & 0.847 & 0.197 \\
Median home value & MHV & Median Home Value (Owner-Occupied Units) & B25077\_001E & 0.769 & 0.189 \\
Homeownership rate & HRE & \% of housing units that are owner-occupied & B25003\_calc\allowbreak\_pctOwnE & 0.248 & 0.216 \\
Households spend & HSE & \% of households spending 50\%+ income on housing & v154\_numerator & 0.125 & 0.175 \\
\end{longtable}

\begin{flushleft}
\footnotesize *Mean and Std are calculated by normalization.
\end{flushleft}

\subsection{Research Process}

\begin{figure}[H]
    \centering
    \includegraphics[width=1\linewidth]{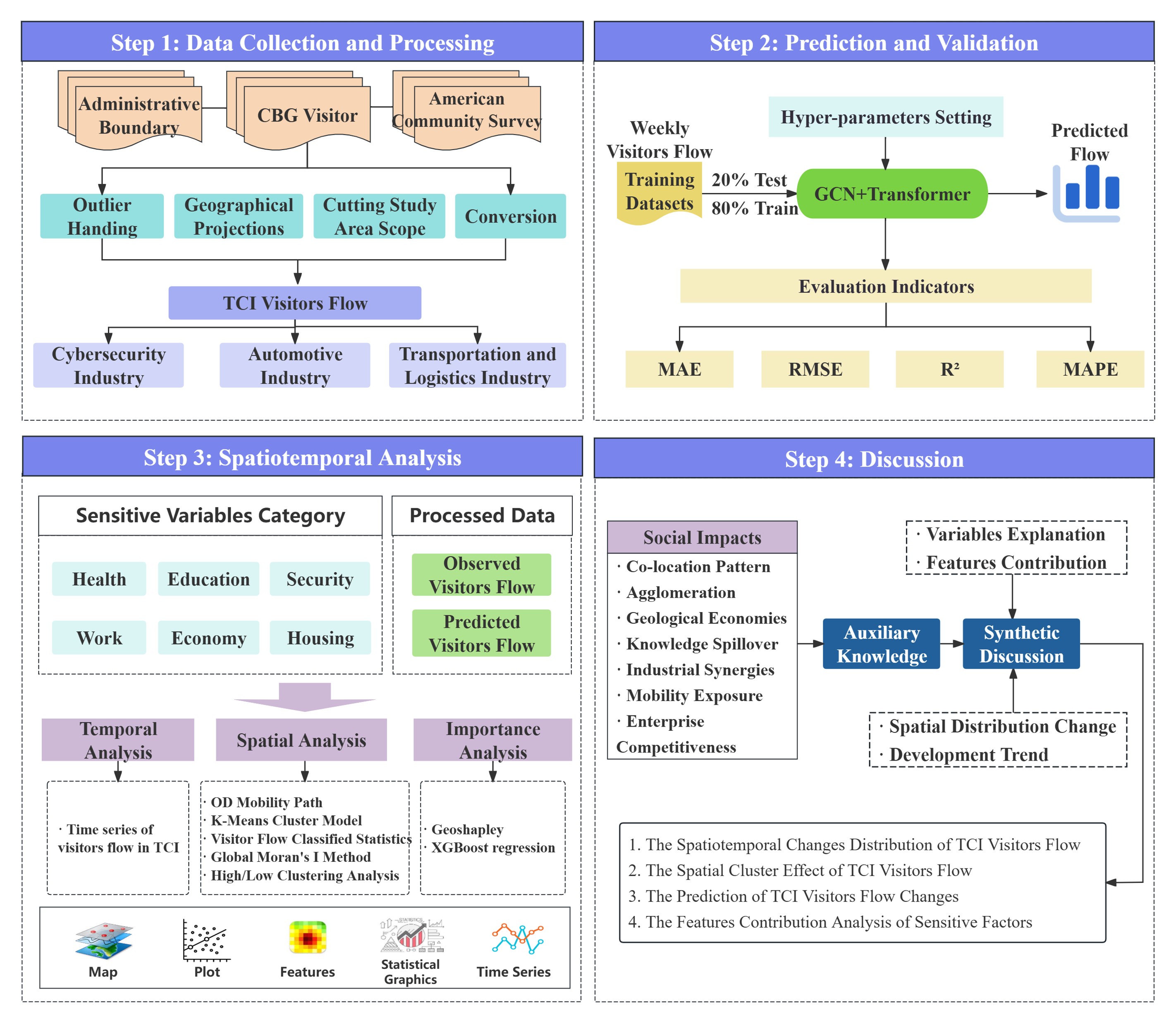}
    \caption{Research framework chart}
    \label{fig:2}
\end{figure}

Fig. \ref{fig:2} presents the proposed step-by-step analytical framework, highlighting our dual focus: the clustering dynamics of TCI-related industries and their relationship with various socioeconomic factors. The process begins with Step 1, where multiple data sources are integrated and pre-processed to align with the research objectives. This step ensures data consistency and prepares the foundation for subsequent analyses. In Step 2, we construct a time-series dataset of visitor flows and develop a high-performance, predictive BiTransGCN model to forecast visitor flows for upcoming periods. To ensure the reliability of the projected visitor flow trends, the model's accuracy is rigorously evaluated using multiple performance metrics. Step 3 involves conducting time-series and spatial analyses on key variables and observed access volume data from 2022. This includes classification statistics, Global Moran’s I method, K-means clustering, and high/low clustering analysis. Specifically, the K-means algorithm is applied to perform a clustering analysis on the visitor flow of the three industries within TCI. Based on the spatial distribution, the network nodes are partitioned into an optimal number of K=6 clusters, with each cluster's centroid representing the average location of the data in that region, and node assignments are iteratively adjusted until the distance between each node and its cluster centroid is optimized \cite{dl1979cluster}. Global Moran’s I is used as a statistical indicator to assess the overall spatial autocorrelation in the visitor flow dataset, describing how variables are clustered or distributed across spatial areas \cite{getis1992analysis}. By quantifying the spatial autocorrelation of visitor flow, the analysis reveals whether TCI-related industries and associated socioeconomic factors are clustered together. The XGBoost and GeoShapley methods are applied for regression testing and to assess the contributions of various factors to the predicted visitor flow trends. A detailed explanation of the methodologies used in Steps 2 and 3 is provided in the following section. Finally, in Step 4, the findings are synthesized to examine how sensitive factors influence access volumes across different industries, exploring their interactions and potential correlations. The discussion extends to broader social and economic impacts, including co-location patterns, industrial clustering, regional economic dynamics, knowledge spillovers, and business competitiveness. 

It is worth noting that this study marks methodological advancement by effectively bridging predictive modeling (Step 2) and exploratory spatiotemporal analysis (Step 3). Unlike traditional prediction methods, which often focus exclusively on either spatial or temporal aspects, the BiTransGCN model uniquely integrates bidirectional temporal relationships and spatial correlations, thus providing a more comprehensive and reliable forecast of visitor flows. Utilizing these prediction results as key variables ensures that our subsequent analysis reflects realistic and dynamically informed scenarios rather than static or retrospective data alone. This approach allows us to uncover nuanced patterns, interactions, and sensitivities among socioeconomic factors and industrial clustering processes over both space and time. 

\section{Methodology}
\subsection{Long-term Flow Prediction Based on Deep Learning}
\begin{figure}[H]
    \centering
    \includegraphics[width=1\linewidth]{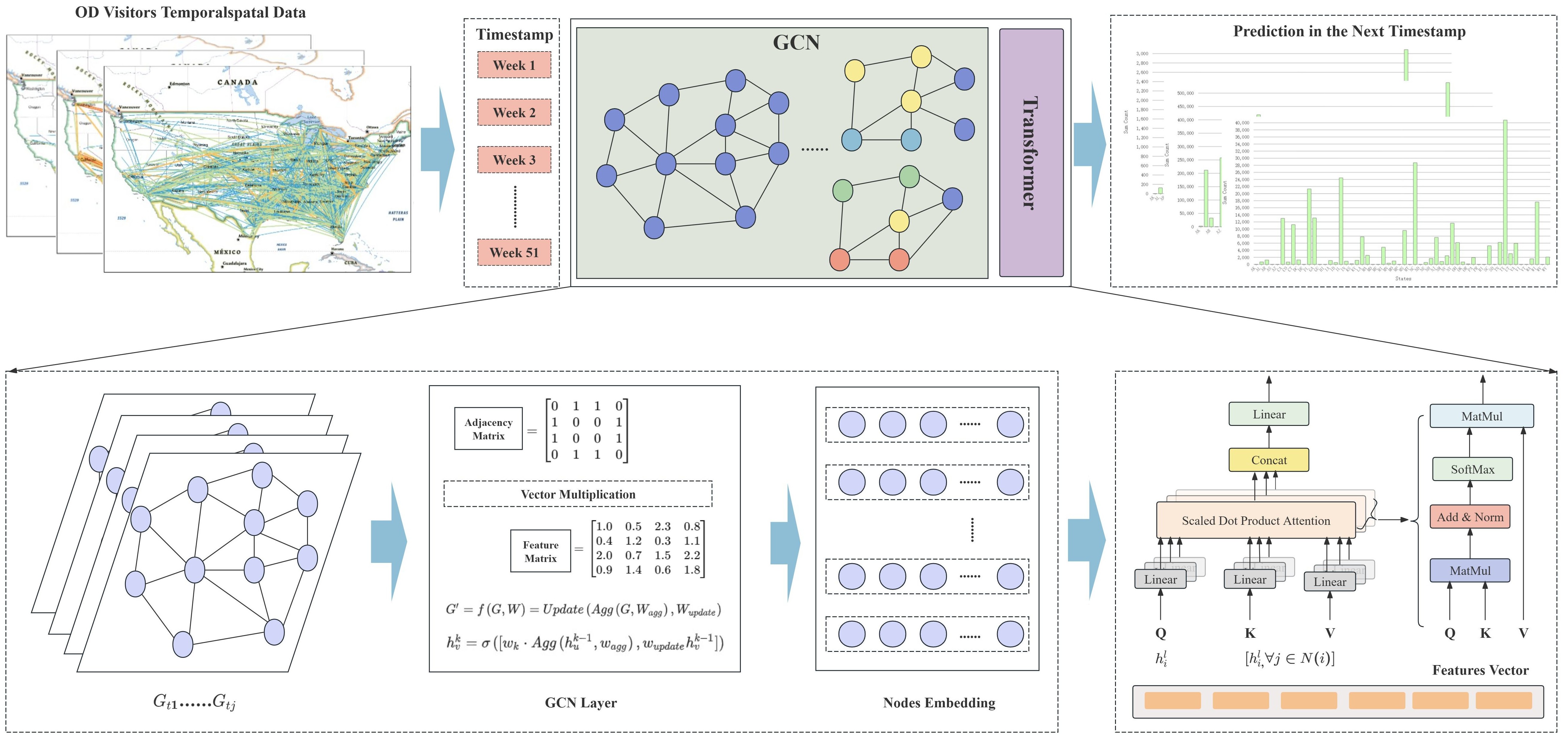}
    \caption{The structure of BiTransGCN model applied for long-term flow prediction}
    \label{fig:3}
\end{figure}

In this study, historical weekly visitor flow data is utilized to project visitor counts for the upcoming period, aligning with the concept of long-term prediction. The proposed BiTransGCN model (Fig. \ref{fig:3}) efficiently handles visitor flow data by seamlessly integrating GCN and Transformer architectures, leveraging the strengths of both to model the complex spatiotemporal patterns inherent in visitor flows. The trained model predicts visitor flows for each spatial unit in the next phase and converts them into change rates to represent TCI dynamics. The process begins with GCN extracting local spatial features from the visitor flow data, representing nodes and their neighboring visitor flow points within the graph structure via graph convolution to produce high-dimensional feature representations. These spatial features are then formatted into distinct time steps to create a time-series input tensor, which is subsequently processed by the Transformer component. Employing the multi-head attention mechanism, the Transformer captures global dependencies across time while preserving the spatial characteristics extracted by GCN. To ensure compatibility with the Transformer, the GCN output is passed through an embedding layer to match the feature dimensions required by the Transformer. Ultimately, the output from the Transformer is mapped to the target space through a decoding layer, enabling the model to integrate both spatial and temporal dependencies effectively. By optimizing both the GCN and transformer components in tandem, BiTransGCN captures detailed spatial information about visitor flows and identifies dynamic temporal patterns, making it highly effective in modeling complex spatio-temporal relationships within visitor flows.

\begin{figure}[H]
    \centering
    \includegraphics[width=1\linewidth]{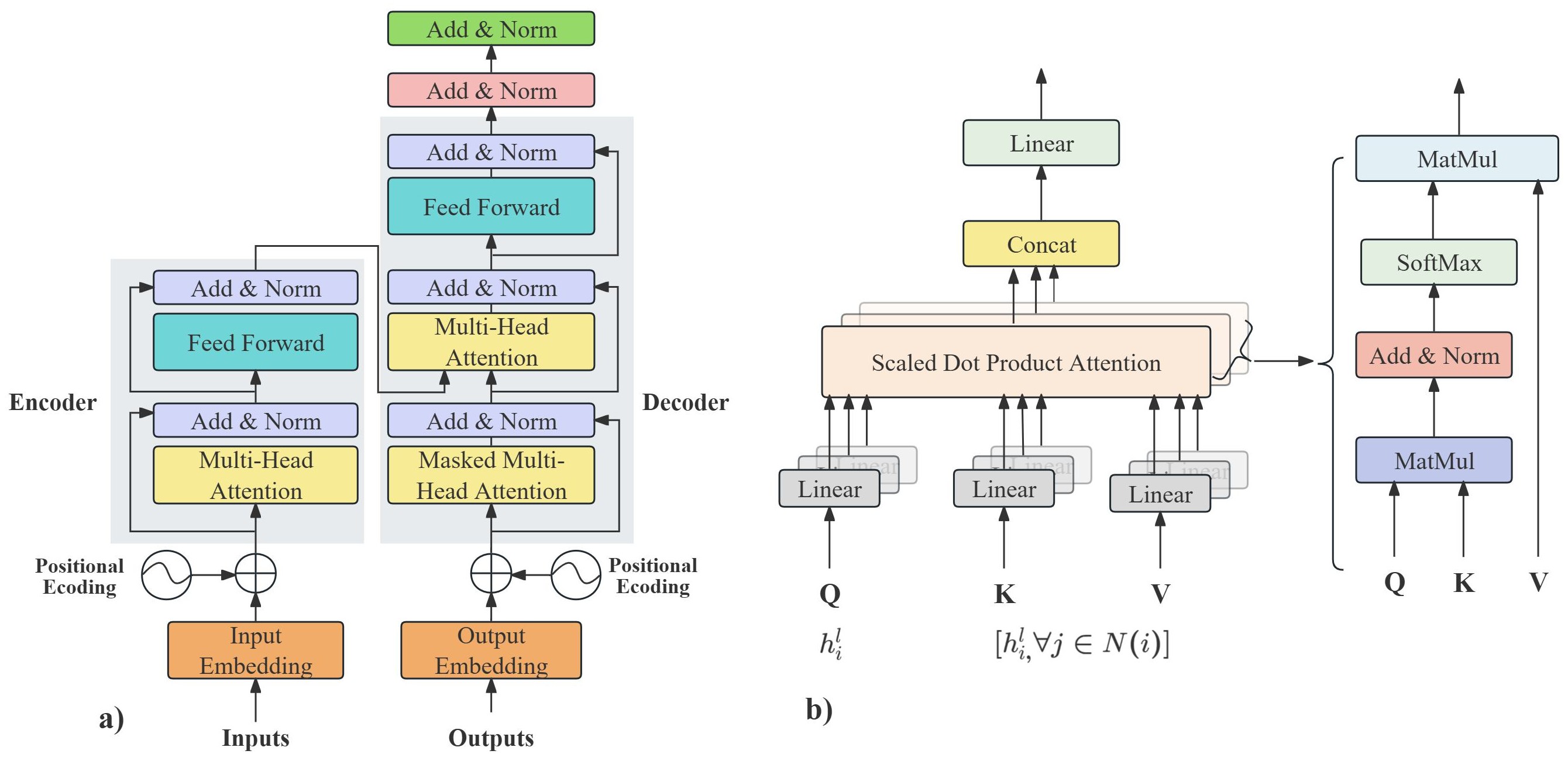}
    \caption{a) Scaled dot-product attention and multi-head self-attention and b) the architecture of transformer}
    \label{fig:4}
\end{figure}

Fig. \ref{fig:4}a illustrates the fundamental architecture of the Transformer model. In our study on visitor flow clustering, weekly visitor-flow data is used as the input sequence. The encoder processes this temporal sequence to extract feature representation vectors that encapsulate inherent temporal dependencies and patterns. Simultaneously, the decoder utilizes both its input sequence and the encoder’s learned feature representations to generate the predicted visitor flows. A key component of the Transformer architecture is its multi-head attention mechanism, which enables the model to attend to multiple time steps in parallel, thereby eliminating the need for sequential state accumulation typical of recurrent models. The generalized attention mechanism \cite{ashish2017attention} is formally expressed as: 

\begin{equation}
    \text{Attention Output} = Att(Q, K, V)
\end{equation}

The variables $Q, K, and V$ denote the query, key, and value derived from the visitor-flow data, respectively. The attention mechanism computes weights by evaluating the interactions between the query and key, which are then applied to the value vectors to obtain a weighted sum that encapsulates temporal dependencies within the visitor flow dataset. In the Transformer architecture, attention is computed using the scaled dot-product mechanism. This can be mathematically expressed as:

\begin{equation}
  \mathrm{Att}(Q, K, V)
  = softmax\!\bigl(\tfrac{QK^{\top}}{\sqrt{d_k}}\bigr)\,V
\end{equation}

$Att$ means Attention, and $d_k$ represents the dimension of key. Fig. \ref{fig:4}b presents the calculation of multi-head attention. The calculation of the $i^th$ attention head can be represented as:

\begin{equation}
    head_i=Att(QW_i^Q,KW_i^K,VW_i^V)
\end{equation}

where $W_i^Q , W_i^K,$ and $W_i^V$ represent the learned linear projection matrices for the query, key, and value vectors in the $i^th$ attention head. Multi-head attention operates by computing attention in parallel across multiple subspaces and then concatenating the results. Let $n$ denote the number of attention heads; the multi-head attention mechanism can be expressed as:

\begin{equation}
    MultiHead(Q,K,V)=concat(head_1,...,head_n)W^L
\end{equation}

In the above formulation, $concat$ denotes the operation of merging outputs from multiple attention heads, each extracting distinct temporal patterns from the weekly visitor-flow data. $W^L$ represents the linear transformation applied to this aggregated visitor flow feature representation. This internal multi-head attention mechanism enables the Transformer to overcome limitations related to non-parallel training and memory capacity that often impede models such as LSTM or GRU in capturing long-term dependencies.

In the GCN backbone, the training dataset $N×C×T$, where $T$ denotes the number of past weeks, $N$ is the number of nodes, and $C$ is the number of features per node. Each node’s features are represented by a feature vector $y_i$, and the complete dataset can be expressed as a collection of these vectors $Y=\{y_1,y_2,y_3,\cdots,y_N\}, y\in R^{N×C×T}$. In the graph structure, key components include nodes, edges, and the global context. Each feature vector is mapped to a graph node, and the entire collection is treated as an unordered set of nodes as $V={v_1,v_2,v_3,\cdots,v_N }$. Connections (edges) are formed between neighboring vectors, representing their relationships. Each edge $e_{ij}$ indicates a directional link from node $v_i$ to node $v_j$, assuming node $v_i$ has $K$ neighbors. These relationships are used in message passing within GCN layers to propagate and aggregate information. The resulting graph is denoted as $G=f(V,E)$, where $E$ is the set of all edges. This framework enables the representation of vector data as a graph structure \cite{kipf2016semi}. The graph is updated as follows:

\begin{equation}
    G^{'}=f(G,W)=Update(Agg(G,W_{agg}),W_{update})
\end{equation}

Through this process, feature vectors of neighboring nodes are aggregated, and a global graph vector is simultaneously obtained. Equations (6) and (7) describe how node features are updated across layers:

\begin{equation}
    h_v^k=\sigma(w_k\cdot Agg(h_u^{k-1},w_{agg}),w_{update}h_v^{k-1})
\end{equation}

\begin{equation}
    Agg=\sum_{u\in{M(v)}}\frac{h_u^{k-1}}{\sqrt{|M(u)||M(v)|}}
\end{equation}

where $h$ denotes the node feature vector, the subscript $u$ indicates a neighboring node, the superscript $k$ refers to the network layer, and $M(v)$ is the set of neighboring nodes for node $v$. $W$ refers to the learnable weight matrices, and $\sigma$ is the activation function $ReLU$, which introduces non-linearity. If the dimensions of the node, edge, and global graph vectors differ, a projection is applied to match the dimensionality of the node vector. In this framework, the global graph vector also acts as a bias term, facilitating complex feature representation across the graph.

\begin{figure}
    \centering
    \includegraphics[width=1\linewidth]{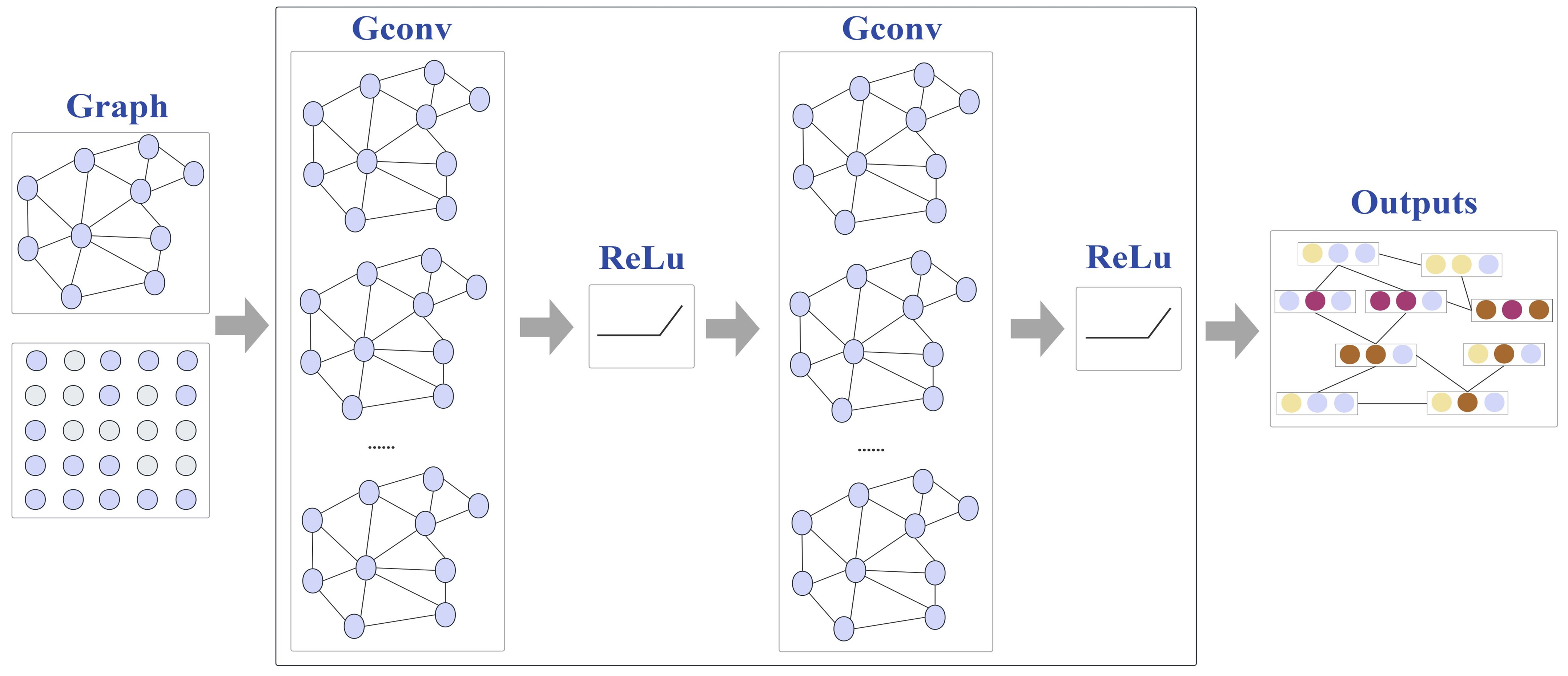}
    \caption{The structure of GCN with multiple graph convolutional layers}
    \label{fig:5}
\end{figure}

In the proposed BiTransGCN model, the structure of GCN consists of multiple graph convolutional layers, as shown in Fig. \ref{fig:5}. The updated weights of GCN constructed will be represented as:

\begin{equation}
    H^{(l+1)}=\sigma(AH^{(l)}W^{(l)})
\end{equation}

$H^{(l)}\in R^{N×F(l)}$:the node representation matrix of the layer $l^th$, initial $H^{(0)}=X. W^{(l)} \in R^{F(l)×F(l+1)}$: the learnable weight matrix of $l^{th}$ layer Normalizing A so that each row sums to one, represented as $D^{-1}A$, where $D$ is the diagonal node degree matrix, resolves this issue. Applying $D^{-1}A$ effectively computes the average of the features of neighboring nodes. However, using symmetric normalization, $D^{-\frac{1}{2}}AD^{\frac{1}{2}}$ introduces more complex dynamics as it moves beyond the simple averaging of neighboring nodes. By combining these approaches, we essentially derive the propagation rule introduced by \cite{kipf2016semi}:

\begin{equation}
    f(H^{(l)},A)=\sigma(\hat{D}^{-\frac{1}{2}}\hat{A}\hat{D}^{-\frac{1}{2}}H^{(l)}W^{(l)})
\end{equation}

With $\hat{A}=A+I$, where $I$ is the identity matrix and $D$ is the degree matrix. $\hat{D}$ is the diagonal node degree matrix of $\hat{A}$, $D_{ij}=\sum_j(A_{ij}+I_{ij})$.

\subsection{Geospatial Factor Importance Analysis Method}
GeoShapley is a spatial data attribution framework designed to quantify and explain the contributions of spatial units or features in spatial decision-making processes. Drawing inspiration from the Shapley Value theory in game theory \cite{harris2021joint}, it provides a systematic and equitable approach to allocating contributions among players in a coalition. In the study, GeoShapley is used to evaluate how a set of social variables and geographic locations impact changes in predicted visitor flows. By ranking the quantified values to represent importance and integrating conclusions from spatial analysis, the framework offers a more scientific interpretation of socio-economic effects. In the algorithm, location features are treated as individual players, allowing for the interpretation of spatial effects such as spatial autocorrelation and spatial heterogeneity \cite{li2024geoshapley}. Following the joint Shapley value, we can calculate $\phi_{GEO}$ as:

\begin{equation}
    \phi_{\mathrm{GEO}} = \sum_{S \subseteq M \setminus \{\mathrm{GEO}\}} \frac{s!(p-s-g)!}{(p-g+1)!} \left( f(S \cup \{\mathrm{GEO}\}) - f(S) \right)
\end{equation}

Here, $p$ represents the total number of players, and we treat the features of spatial units as the "players" in the model. The set ${j}$ includes all possible combinations of players except for j, while $S$ is a subset of ${j}$ containing s players. $f(S)$ denotes the outcome based on the players in $S$, and $f(S\cup\{j\})
$ represents the outcome when player $i$ is added to $S$. Additionally, $GEO$ is a set of location features with a size of $g$.
The interaction effect between location and features can be determined using the Shapley interaction value, which is defined as follows:
\begin{equation}
\begin{aligned}
    \phi_{(\mathrm{GEO},\,i)} &= \sum_{S \subseteq M \setminus \{\mathrm{GEO}\}} \frac{s! (p-s-g)!}{(p-g+1)!} \, \Delta(\mathrm{GEO},i) \\
    \Delta(\mathrm{GEO},i) &= f(S \cup \{\mathrm{GEO},i\}) - f(S \cup \{\mathrm{GEO}\}) - f(S \cup \{i\}) + f(S)
\end{aligned}
\end{equation}

Using a similar estimation process as kernel SHAP, GeoShapley values can be calculated for each individual observation and then averaged across the background dataset. The final GeoShapley output consists of four components that collectively sum up to the model's prediction:

\begin{equation}
    \hat{y} = \phi_0 + \phi_{\mathrm{GEO}} + \sum_{j=1}^{p} \phi_j + \sum_{j=1}^{p} \phi_{(\mathrm{GEO},\,j)}
\end{equation}

Where $\phi_0$ represents a constant base value, which is the average prediction based on background data and acts as the global intercept. $\phi_GEO$ is a vector of size $n$ that quantifies the intrinsic location effect in the model. $\phi_j$ is a vector of size $n$ for each non-location feature $j$, capturing the location-invariant effect in the model. $\phi_{(GEO,j)}$ is a vector of size $n$ for each non-location feature $j$, representing the spatially varying interaction effect in the model. If no spatial effects are in the model, both $\phi_GEO$ and $\phi_{(GEO,j)}$ will be zero.

\section{Results and Discussion}
\subsection{Spatiotemporal Changes of Visitor Flows in TCI-related Industries}
\begin{figure}[H]
    \centering
    \begin{subfigure}{\textwidth}
        \centering
        \includegraphics[width=0.8\textwidth]{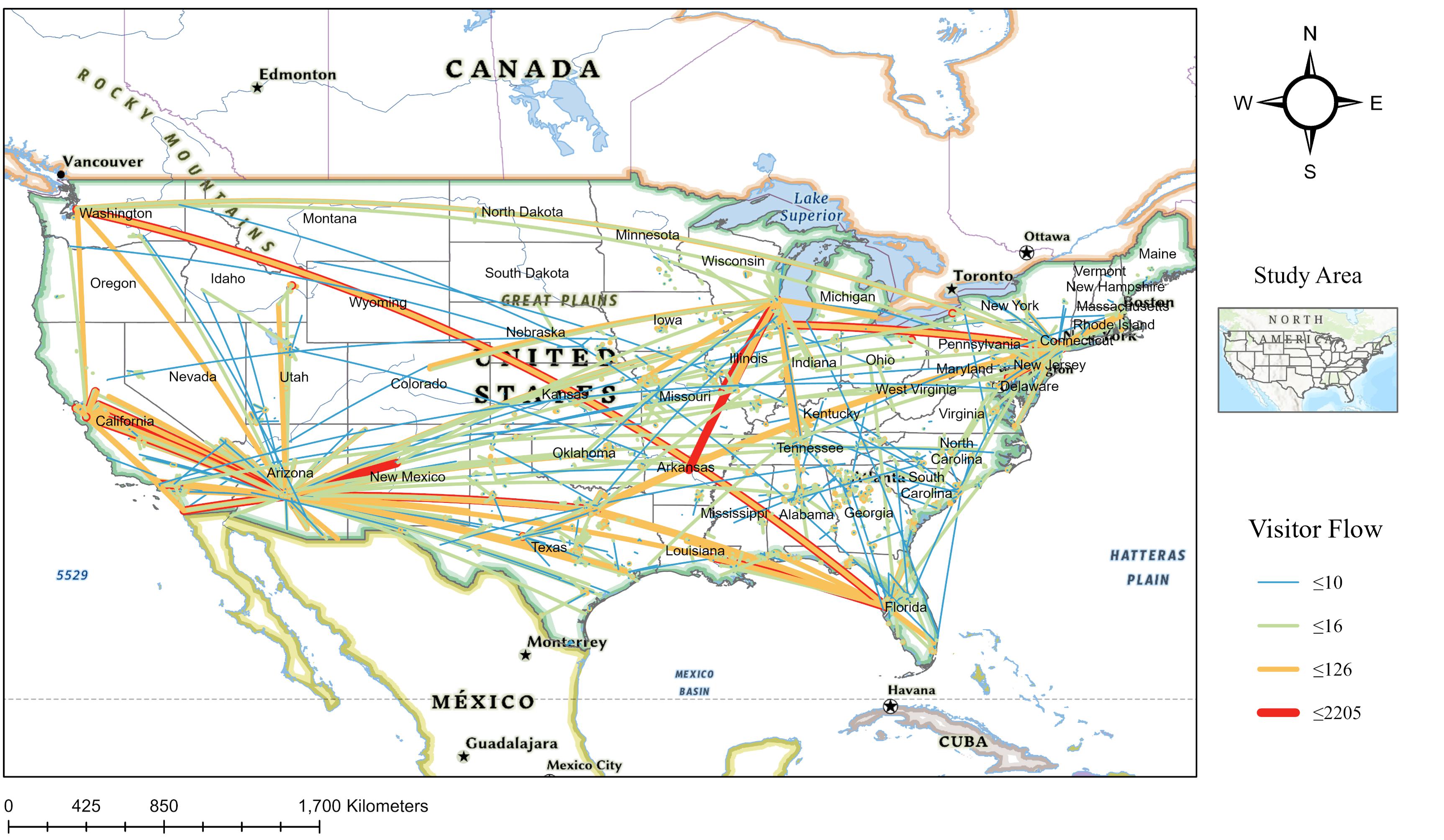}
        \caption{The OD map of automotive industry}
        \label{fig:6a}
    \end{subfigure}
    
    \vspace{0.5em} 

    \begin{subfigure}{\textwidth}
        \centering
        \includegraphics[width=0.8\textwidth]{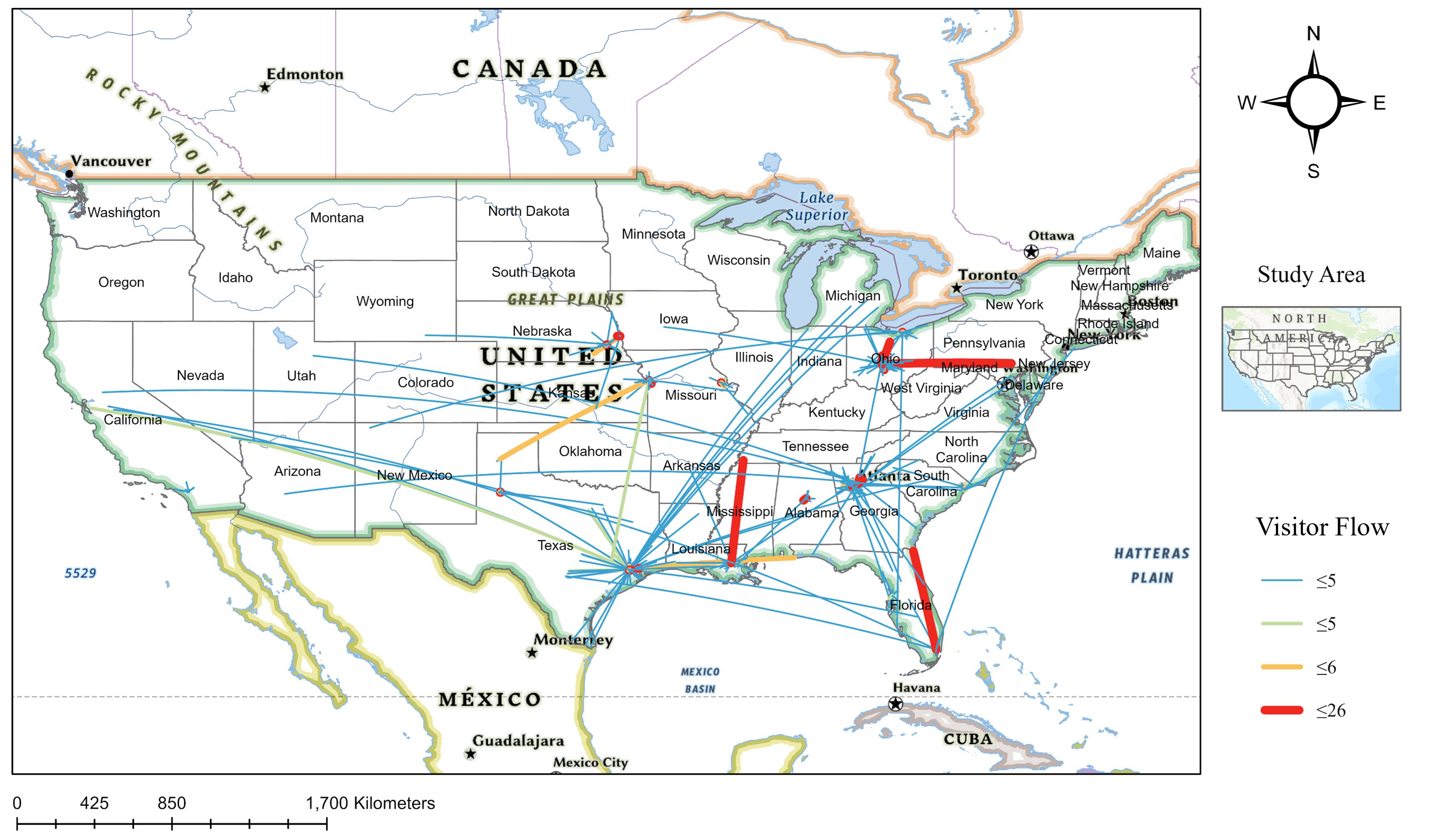}
        \caption{The OD map of cybersecurity industry}
        \label{fig:6b}
    \end{subfigure}
    
    \vspace{0.5em} 

    \begin{subfigure}{\textwidth}
        \centering
        \includegraphics[width=0.8\textwidth]{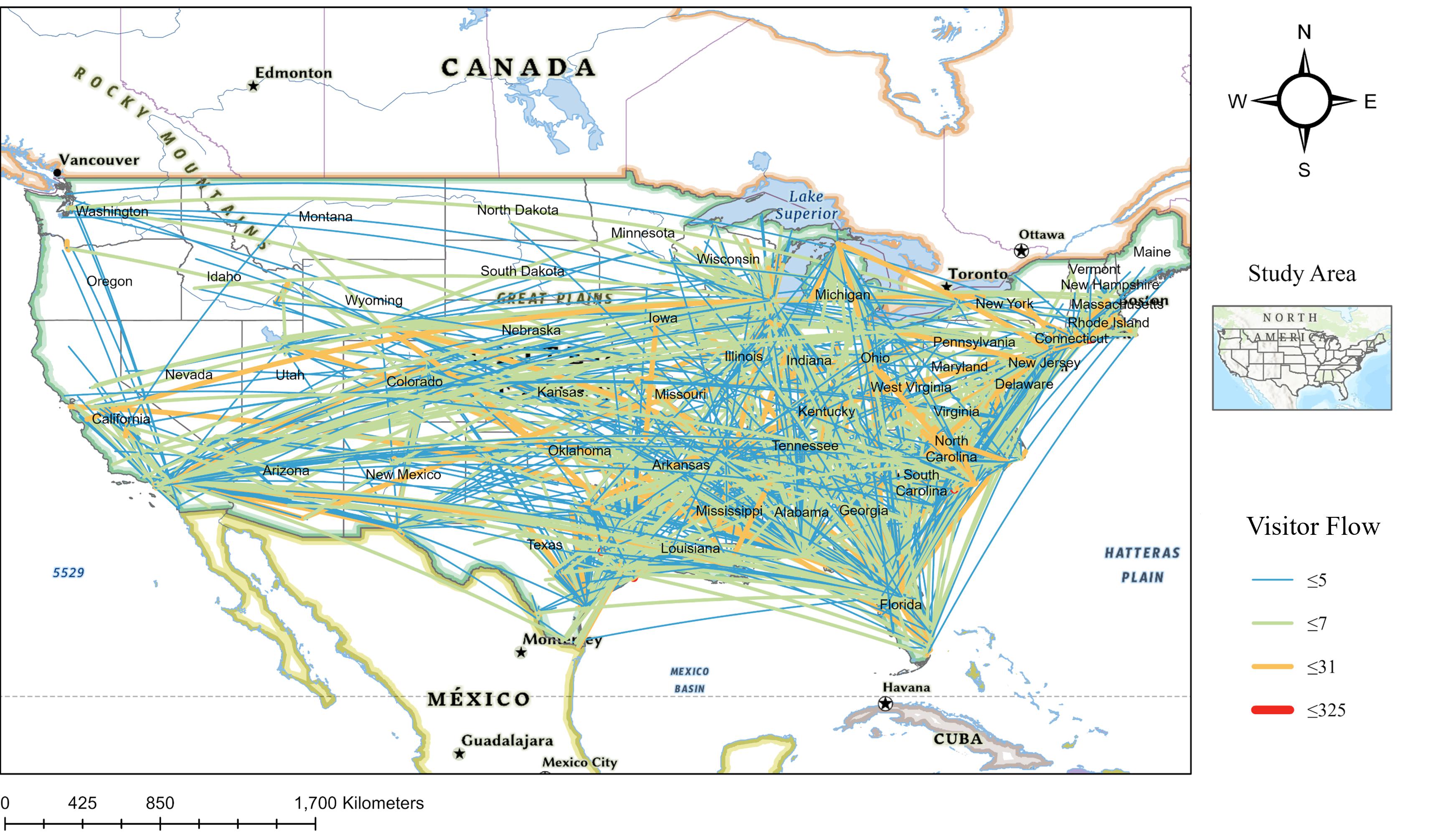}
        \caption{The OD map of transportation and logistics industry}
        \label{fig:6c}
    \end{subfigure}

    \caption{The OD map of three industries in TCI visitors}
    \label{fig:6}
\end{figure}

The geographic distribution of visitor coverage across the three TCI-related industries in the U.S. reveals striking differences in spatial reach and travel patterns. Based on the density of OD lines, the transportation and logistics industry (Fig. \ref{fig:6c}) exhibits the widest coverage, with high-frequency visits (yellow and red lines) primarily concentrated in the northern and central regions. 

The automotive industry (Fig. \ref{fig:6a}) follows in terms of coverage area but surpasses the other two industries in visit volume. Travel routes in this sector are notably longer, often spanning half the country, with major activity hubs in California, Washington, Arizona, and Wisconsin. In contrast, the cybersecurity industry (Fig. \ref{fig:6b}) has a more localized footprint, with high-frequency visits concentrated in Louisiana, Florida, Tennessee, New York, and Ohio, and significantly shorter travel distances. 

These patterns highlight how each industry is organized geographically and functionally. The automotive industry, for example, was historically centered in Michigan but has expanded over time \cite{klier2005determinants}. Today, Pennsylvania and New York have become important hubs in the Northeast, while Kentucky, Tennessee, and Texas have seen significant growth in the South. Meanwhile, Arizona’s role as a semiconductor production hub underscores how regional specialization influences visitor travel paths and coverage patterns. These findings illustrate how industrial geography, supply chain dynamics, and regional economic shifts shape industry mobility trends.

\begin{figure}[H]
    \centering
    \begin{subfigure}{\textwidth}
        \centering
        \includegraphics[width=0.88\textwidth]{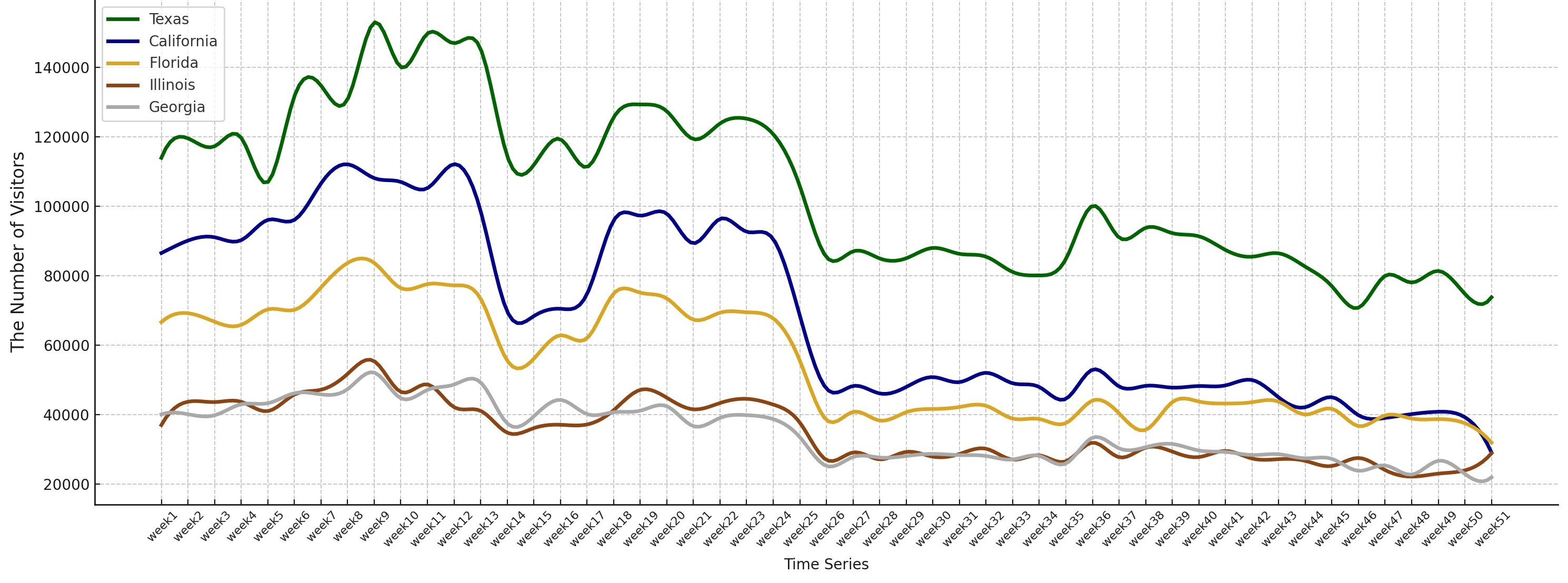}
        \caption{Time series of top 5 states by visitor flow in the TCI over time}
        \label{fig:7a}
    \end{subfigure}
    
    \vspace{1em} 

    \begin{subfigure}{\textwidth}
        \centering
        \includegraphics[width=0.88\textwidth]{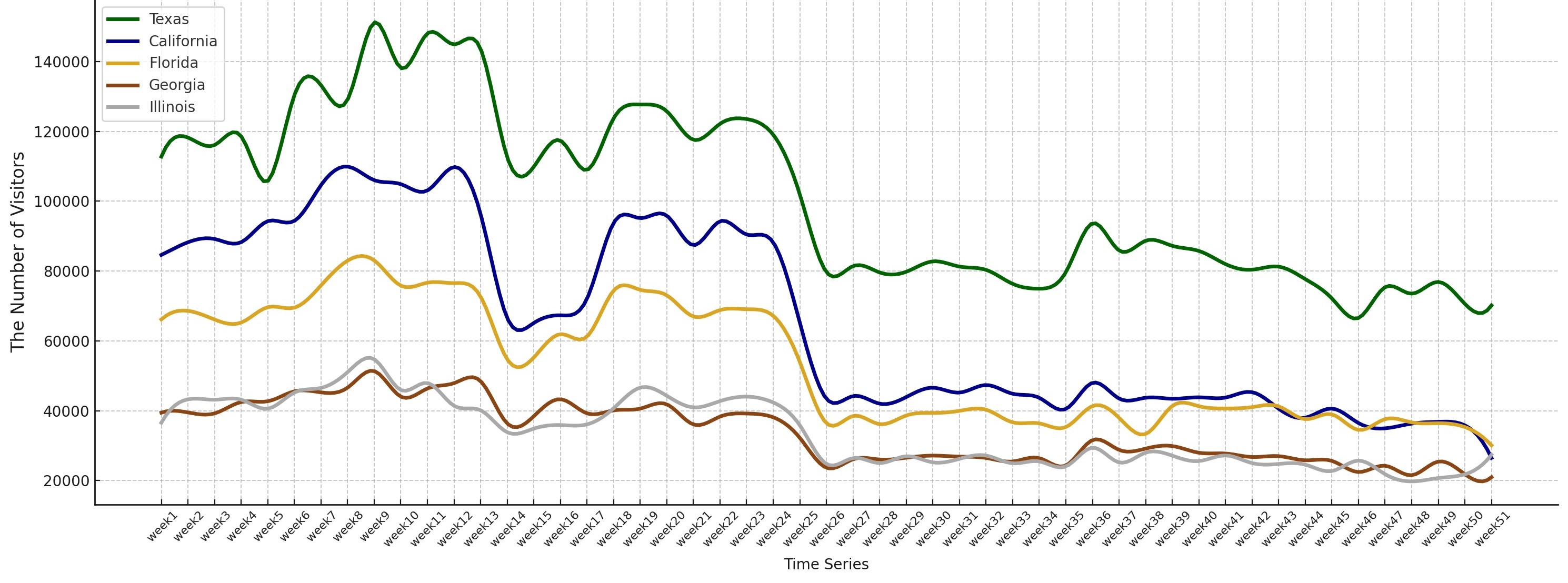}
        \caption{Time series of top 5 states by visitor flow in the automotive industry over time}
        \label{fig:7b}
    \end{subfigure}
    
    \vspace{1em} 

    \begin{subfigure}{\textwidth}
        \centering
        \includegraphics[width=0.88\textwidth]{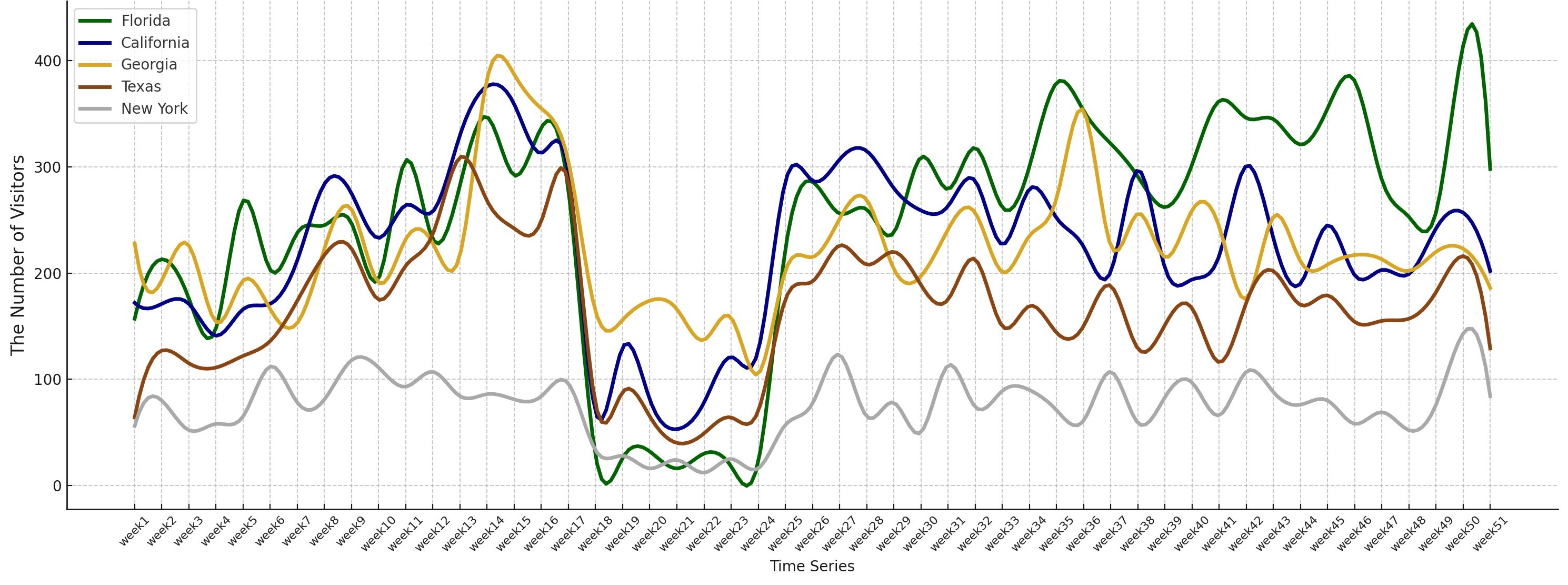}
        \caption{Time series of top 5 states by visitor flow in the cybersecurity industry over time}
        \label{fig:7c}
    \end{subfigure}

        \vspace{1em} 

    \begin{subfigure}{\textwidth}
        \centering
        \includegraphics[width=0.88\textwidth]{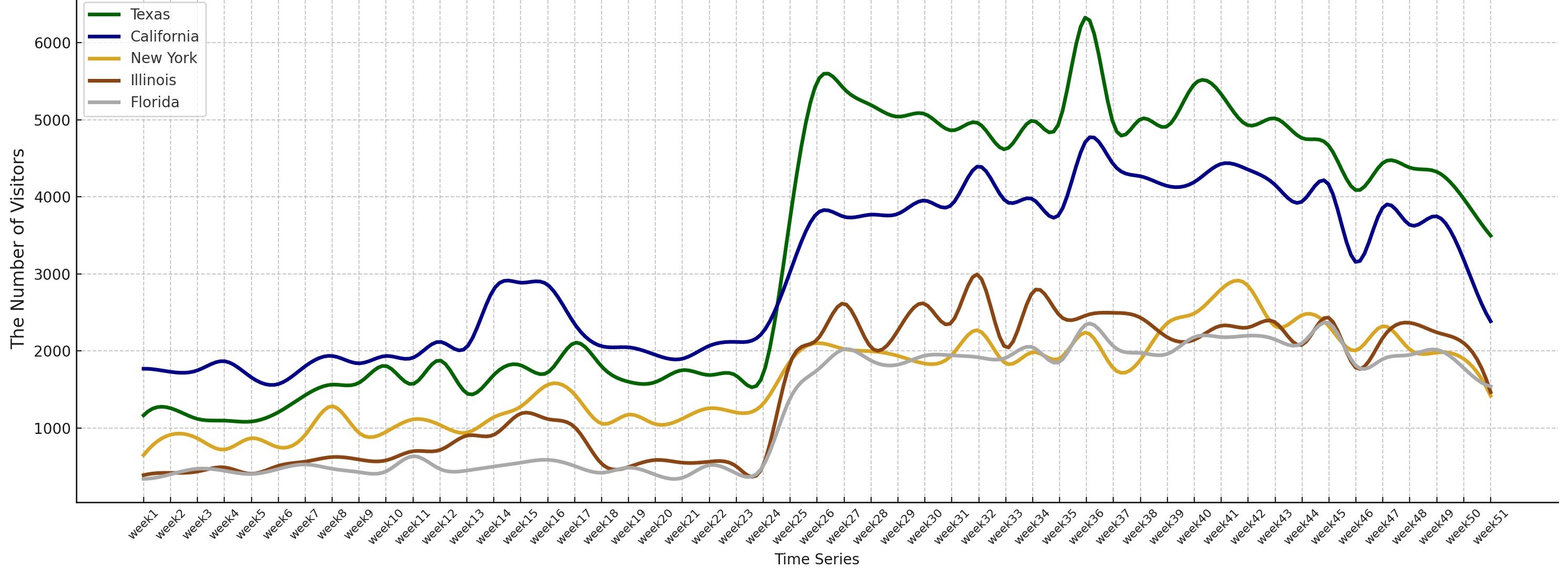}
        \caption{Time series of top 5 states by visitor flow in the transportation and logistics industry over time}
        \label{fig:7d}
    \end{subfigure}

    \caption{Time series of visitor flows in three industries in TCI}
    \label{fig:7}
\end{figure}

From a temporal perspective, the automotive industry consistently recorded the highest visitor flows across all 51 weeks of 2022, outperforming the other two industries (Fig. \ref{fig:7}). To further explore the evolution of visitor flows at the state level, we focused on the top five states. Although each industry’s pattern was distinct, the visitor flow trends among states within the same industry were almost identical. For TCI, which combines data from all three industries (Fig. \ref{fig:7a}), Texas maintained the highest visitor flow throughout the study period, followed by California. In the automotive industry (Fig. \ref{fig:7b}), Texas again stood out with a notably higher visitor flow than any other state, peaking in Week 9. Georgia and Illinois exhibited similar, often alternating, visitor flows. Overall, the automotive industry dominated TCI because it contributed the largest volume of visits. In the cybersecurity industry (Fig. \ref{fig:7c}), Florida consistently led in visitor flow, though its advantage was not pronounced until after Week 30, while New York ranked last among the top five states. Finally, in the transportation and logistics industry (Fig. \ref{fig:7d}), the trends across states were broadly consistent. Texas initially trailed California but moved into first place in Week 25 and remained the leader, while New York, Illinois, and Florida maintained closely aligned visitor flows throughout the study period.

\subsection{Spatial Clustering Feature Analysis}
\begin{figure}[H]
    \centering
    \includegraphics[width=1\textwidth]{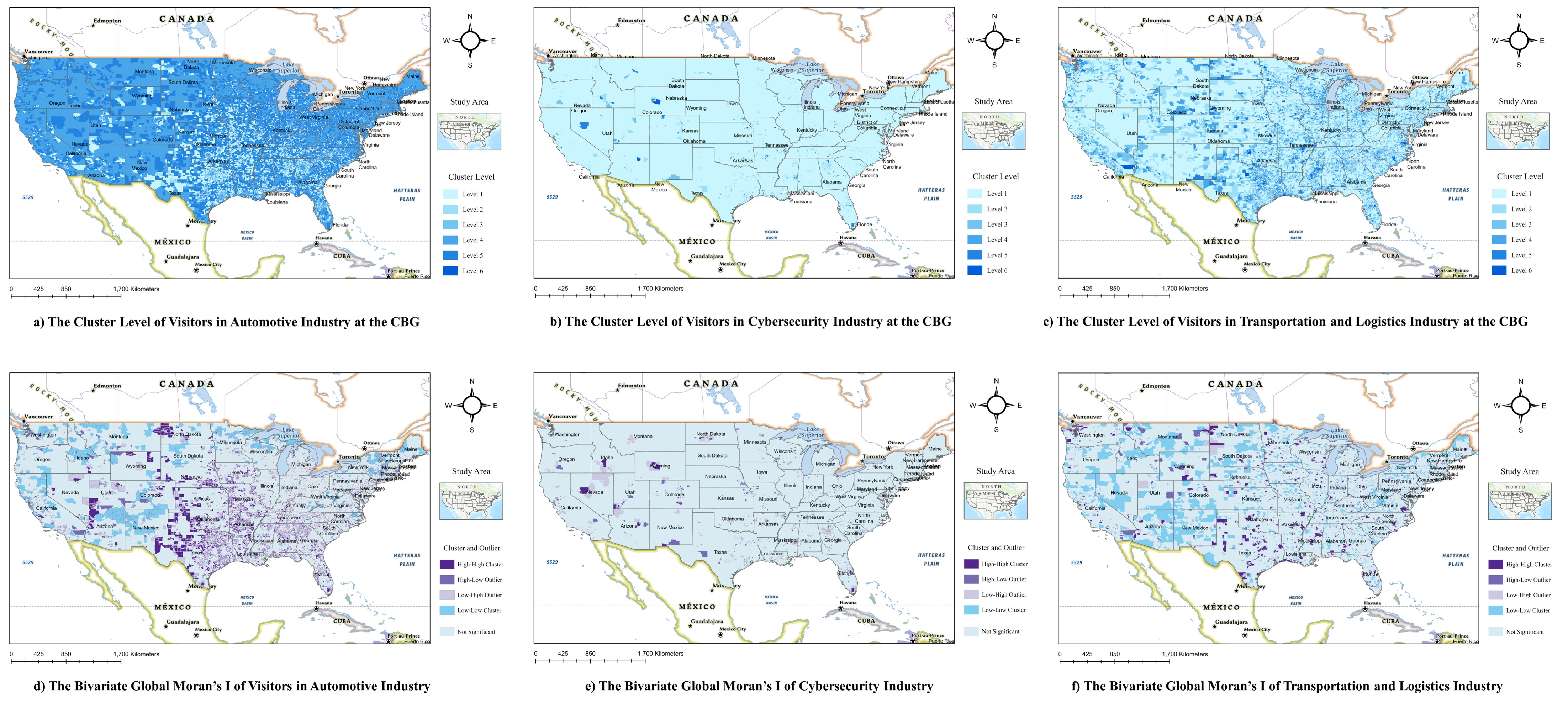}
    \caption{The cluster level of visitor flows in TCI at the CBG (a-c), and the bivariate global Moran’s I maps (d-f)}
    \label{fig:8}
\end{figure}

To explore the spatial distribution of visitor flows, we applied a multivariate clustering approach to identify spatial clustering levels and categorize the relevant statistics. By normalized processing, the classification ranges for cluster levels from level 1 to level 6 are respectively defined as follows: 0-0.17, 0.18-0.33, 0.34-0.50, 0.51-0.67, 0.68-0.83, and 0.84-1.0. Fig. \ref{fig:8}(a-c) presents these clustering levels for TCI at the CBG level, showing where visitors originate and highlighting their spatial clustering patterns. Among the three industries, the automotive sector exhibits substantially higher cluster levels and coverage areas than the others. In particular, the western and northern regions of the U.S. generally show more pronounced clustering than the southern and eastern regions.

In contrast, the cybersecurity industry shows mainly lower cluster levels (levels 1 and 2), except for small, highly concentrated areas in Oregon and Colorado. Meanwhile, the transportation and logistics industry features a more uniform distribution of visitor flows: nearly every state includes all clustering levels, except for a few sparsely populated areas.


\begin{figure}[H]
    \centering
    \includegraphics[width=1\textwidth]{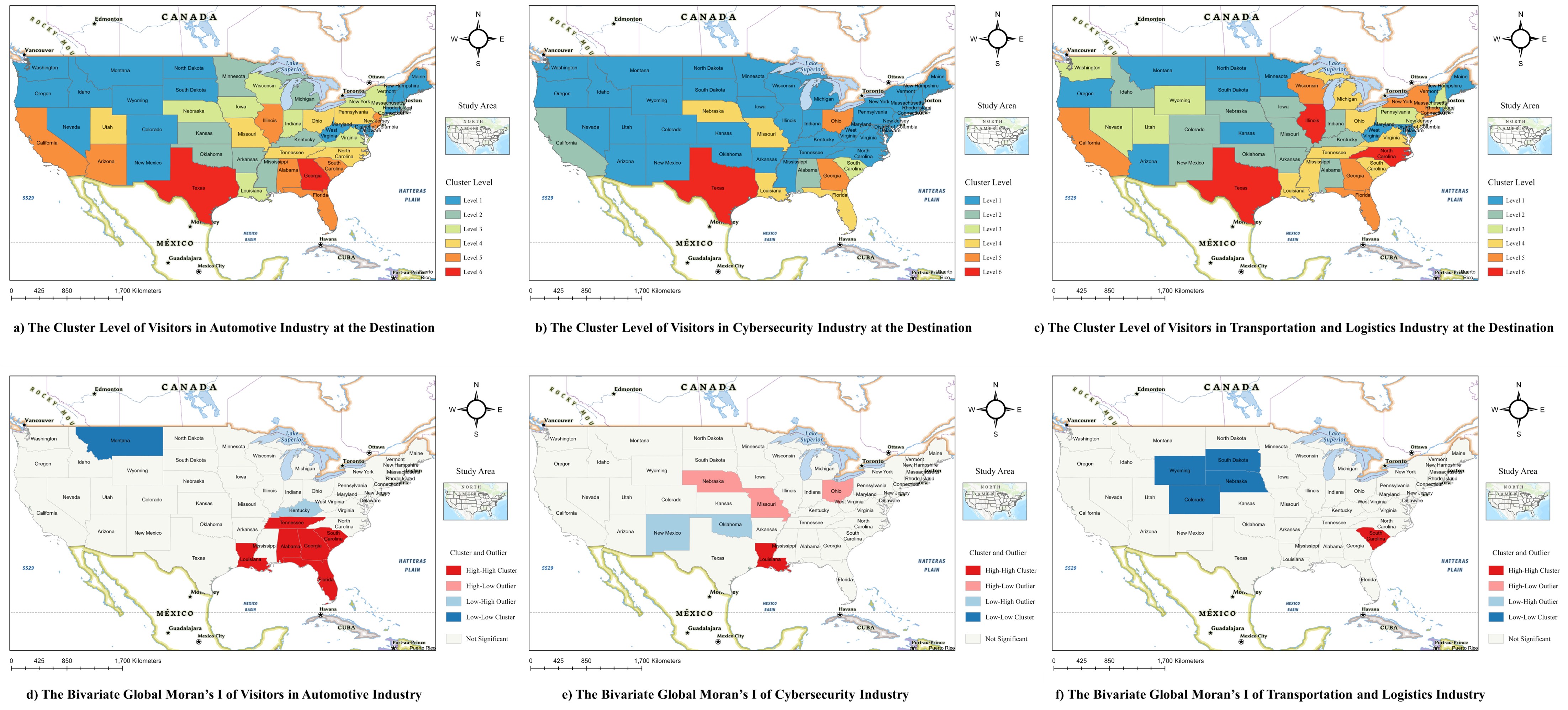}
    \caption{The cluster level of visitor flows in TCI at the Destination (a-c) and the bivariate global Moran’s I maps (d-f)}
    \label{fig:9}
\end{figure}

Global and local bivariate Moran’s I provide a means to uncover spatial correlations and dependencies among neighboring regions. As shown in Fig. \ref{fig:9}(d-f), this analysis reveals distinct spatial autocorrelation patterns for TCI visitors. Within the automotive industry, high-high clusters are largely concentrated in the central-southern and eastern U.S. These areas are geographically adjacent, forming a pronounced spatial clustering effect. The reasons include established manufacturing clusters, robust supply chain networks, and substantial market demand. Low-low or low-high clusters are distributed across parts of the western and northern U.S, suggesting that these regions exhibit either relatively low visitation in the automotive sector or a differentiated spatial distribution pattern compared to neighboring areas. In contrast, high-high cybersecurity clusters appear primarily in Idaho, Nevada, Wyoming, New Mexico, and parts of Texas. These areas may benefit from the presence of data centers, technological innovation hubs, and companies associated with the defense industry, which attract a larger volume of visitors. In the transportation and logistics industry, high-high clusters are observed in most states but are especially prominent in the Midwest. This region typically features well-developed transportation networks, rail hubs, and highway systems that facilitate the growth of logistics industries and draw more visitors from related fields. Compared to the other industries, the sector demonstrates a broader spatial distribution, reflecting a strong level of visitation nationwide, yet its core clusters remain relatively concentrated in states with marked advantages in transportation infrastructure.

Drawing on the preceding analyses, these distinct geographic clustering patterns have substantial implications for industry development and strategic decision-making. The regional disparities underscored by these clusters can inform targeted policies and interventions related to industrial layout, infrastructure investment, and economic expansion. Each cluster type reflects a unique pattern of high or low visitor concentration, shaped by factors such as supply chain networks, educational attainment, labor market dynamics, and workplace conditions. 


Fig. \ref{fig:9} illustrates how visitor clustering levels vary across three related industries significantly. In the automotive sector (Fig. \ref{fig:9}a), Texas and Georgia stand out with the highest clustering levels, followed by the Southwest (California and Arizona) and the Southeast (Alabama, Florida, and South Carolina). Illinois is the only major cluster in the central region. Notably, these clusters differ in character. Based on the global Moran’s I index (Fig. \ref{fig:9}d), the Southwest exhibits a strong high-high clustering pattern, while areas around Montana present a contrasting low-low pattern. Other regions show no statistically significant spatial clustering, suggesting that their clustering patterns remain geographically independent.

In the cybersecurity industry (Fig. \ref{fig:9}b), Texas emerges as the sole region with the highest clustering level (Level 6), followed by Georgia and Ohio. Most other areas do not exhibit significant clustering. According to the global Moran’s I results (Fig. \ref{fig:9}e), Louisiana has a notable high-high clustering pattern, while a few central states display high-low and low-high clusters. Statistically, the rest of the country does not demonstrate meaningful spatial clustering. The transportation and logistics industry (Fig. \ref{fig:9}c) similarly identifies Texas with the highest clustering level (Level 6), with Illinois and North Carolina showing comparably high levels. Overall, the eastern and southern United States display stronger clustering than the western and northern regions. South Carolina exhibits a distinct high-high clustering pattern (Fig. \ref{fig:9}f), whereas Colorado, Wyoming, South Dakota, and Nebraska in the central region reveal a clear low-low pattern.

The geographic distribution of these industries is crucial for clustering professionals, promoting technological exchange, and enhancing labor mobility. The automotive sector, once centered in Michigan and the Northeast, has steadily expanded southward through Tennessee into northern Mississippi, Alabama, and Texas, with Georgia linking the Midwestern industrial core to East Coast clusters \cite{klier2005determinants}. In the southern automotive corridor, foreign supplier plants outnumber domestic ones by 2.4 times, reflecting the strong interplay between customer locations and suppliers \cite{florida1994agglomeration}.

Texas, a major maritime and rail hub, moves roughly 70\% of North America’s finished vehicles by rail—a figure that continues to rise. In 2014, cross-border import trade reached USD 185 billion, underscoring Texas’s vital rail network \cite{harrison2015intra}. The state’s automotive and related manufacturing sectors also rely heavily on Intra-Industry Trade (IIT) for competitiveness \cite{galvin2015recent}, leading the nation in transportation equipment and machinery trade. High freight volumes along the IH-35 corridor, which often crosses the U.S.-Mexico border, further intensify Texas’s clustering of visitor flows.

Since TCI thrives on technology, accessible transportation is essential. Companies that relocate away from established clusters risk losing benefits such as knowledge spillovers, workforce mobility, and peer interactions. Strengthening security with supply chain stakeholders is vital for overall security, while growing cross-regional partnerships stimulate the flow of industry professionals and reinforce the importance of strategic clustering for competitiveness.

\subsection{Prediction of Visitor Flow Changes of TCI-related Industries}
\begin{figure}[H]
    \centering
    \includegraphics[width=1\linewidth]{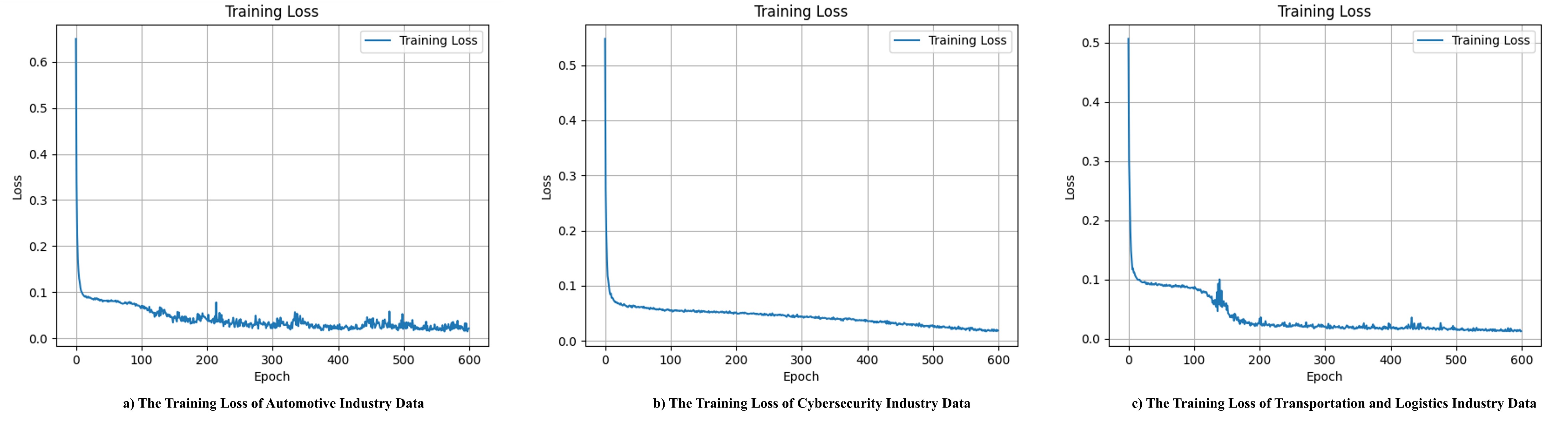}
    \caption{The training loss of BiTransGCN model in TCI visitor flow of prediction}
    \label{fig:10}
\end{figure}

The proposed BiTransGCN model was used to forecast state-level visitor flows for the next week for three types of industries. The data from 2022 was collected and processed, and a 4:1 split was applied to create training and testing datasets. Through extensive testing, the best parameter settings were determined: dropout rate = 0.05, learning rate = 0.0001, hidden layer size = 128, weight decay = 1e-5, and 600 training epochs. As shown in Fig. \ref{fig:10}, the model's training loss is mean squared error (MSE), which decreased and converged smoothly, ensuring optimal and effective results. This setup avoided common issues such as overfitting, underfitting, or getting stuck in suboptimal solutions. Standard evaluation metrics for deep learning models, including mean absolute error (MAE), root mean squared error (RMSE), coefficient of determination (R²), and mean absolute percentage error (MAPE) were used to assess prediction accuracy (Table \ref{tab:4}). The results showed that the model performed best for the automotive industry, followed by the transportation and logistics industry, while predictions for the cybersecurity industry were also accurate. It indicates that the predicted TCI data is reliable, and the insights derived from these forecasts are credible and theoretically persuasive. 

\begin{figure}[H]
    \centering
    \includegraphics[width=1\linewidth]{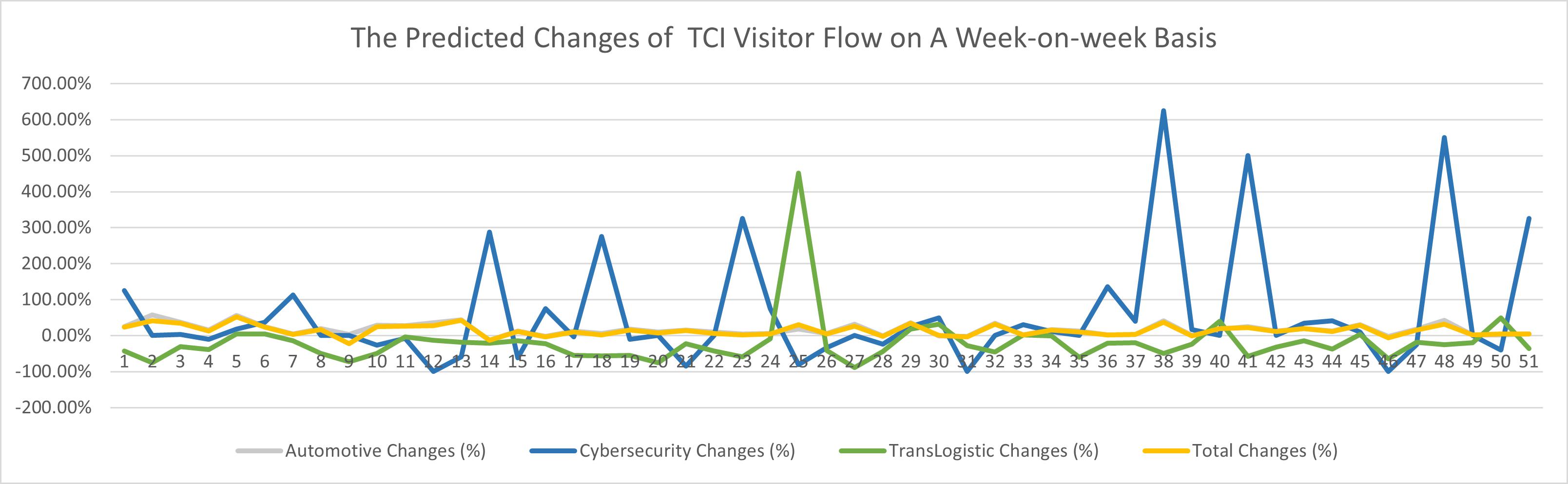}
    \caption{The prediction changes of TCI visitor flows on a week-on-week basis}
    \label{fig:11}
\end{figure}

Based on the predicted visitor flows for the three industries, we calculated the rate of change for each industry and TCI overall in the next period using the week-on-week concept. A statistical chart of the rate of change for each state in the U.S. was plotted (Fig. \ref{fig:11}). Clearly, TCI, with a yellow line representing the overall industry trend, shows growth in all 51 states, with an average rate of change of approximately 14.16\%. The growth trend of the automotive industry closely aligns with TCI, with an average rate of change of 16.72\%. In comparison, the cybersecurity industry exhibits more volatility, with some states even experiencing negative growth. However, the proportion of states with growth far exceeds those with negative growth, resulting in an average rate of change of 58.14\%. Notably, Oregon, South Carolina, and Washington show the most significant growth, each exceeding 500\%. On the other hand, the transportation and logistics industry has more states with negative growth than positive growth, resulting in an average rate of change of approximately -18.77\%.


\begin{table}[htbp]
\centering
\renewcommand{\arraystretch}{1.3} 
\setlength{\tabcolsep}{12pt} 
\caption{The accuracy evaluation of visitor flow predictions for the three categories in TCI}
\label{tab:4}
\begin{tabular}{lcccc}
\toprule
\textbf{TCI Categories/ Evaluation Indicators} & \textbf{MAE} & \textbf{RMSE} & \textbf{R\textsuperscript{2}} & \textbf{MAPE} \\
\midrule
Automotive Industry & 0.071 & 0.136 & 0.927 & 18.86\% \\
Cybersecurity Industry & 0.506 & 0.447 & 0.867 & 24.40\% \\
Transportation and Logistics Industry & 0.183 & 0.395 & 0.940 & 21.92\% \\
\bottomrule
\end{tabular}
\end{table}

\subsection{Spatial Effects and Features Contribution Analysis of Socioeconomic Factors}
\begin{figure}[H]
    \centering
    \includegraphics[width=1\textwidth]{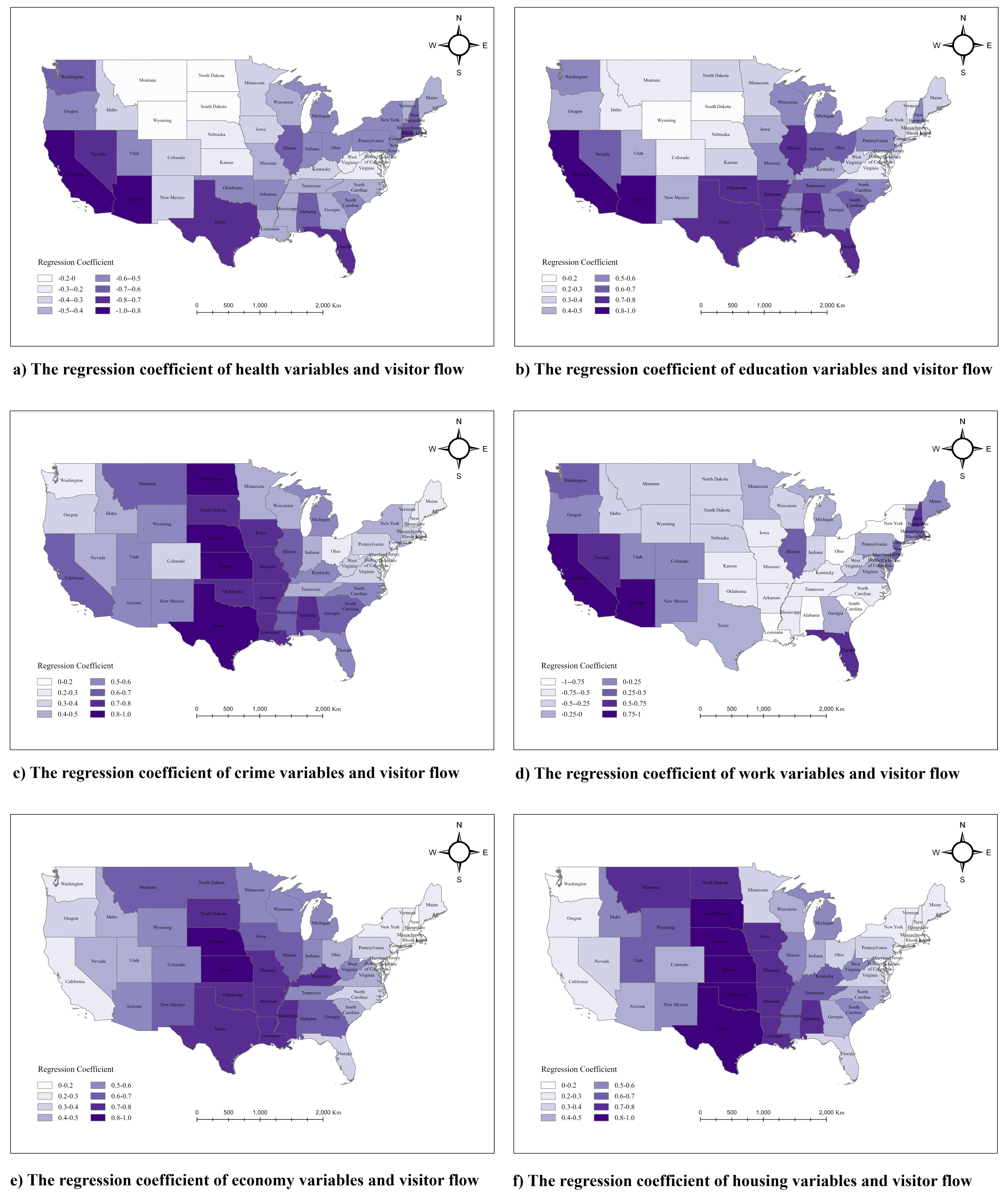}
    \caption{The regression coefficient maps (a-f) between 6 categories of variables and visitor flows}
    \label{fig:12}
\end{figure}

Six socioeconomic factors are used as independent variables. The previously predicted rates of change in visitation volumes for each industry are the dependent variable. Since the sparsity-aware algorithm for parallel tree learning enhances computational performance in XGBoost \cite{chen2016xgboost}. Beyond that, the model’s residuals exhibit strong normality with minimal spatial autocorrelation \cite{li2024geoshapley}. It robustly reflects the strength and distribution of the correlations between socioeconomic factors and TCI visitor flow variables. Therefore, it is reasonable that we employed the XGBoost model to calculate GeoShapley values and interpret the feature contributions of the variables. Specifically, the regression relationship between the independent (each social variable) and dependent (TCI visitor flow variable) was used as an example to analyze their potential connections (Fig. \ref{fig:12} a-f), which can overall reflect the strength of the correlation and its distribution. The social variables consist of health variables, education variables, crime variables, work variables, economy variables, and housing variables, with each type representing a weighted calculation of three sub-variables from their respective categories.

In the experiment, the XGBoost model sets up hyperparameter tuning using Hyperopt. The learning rate was configured within a range of 0.001 to 0.1, the maximum depth was selected from 3, 5, 7, or 10, and the number of trees ranged from 50 to 500 in increments of 50. Both the subsample ratio and the feature sampling ratio were adjusted within a range of 0.5 to 1.0. The model was evaluated using RMSE as the performance metric and was set to run for a maximum of 500 iterations.

\begin{figure}[H]
    \centering
    \includegraphics[width=0.84\textwidth]{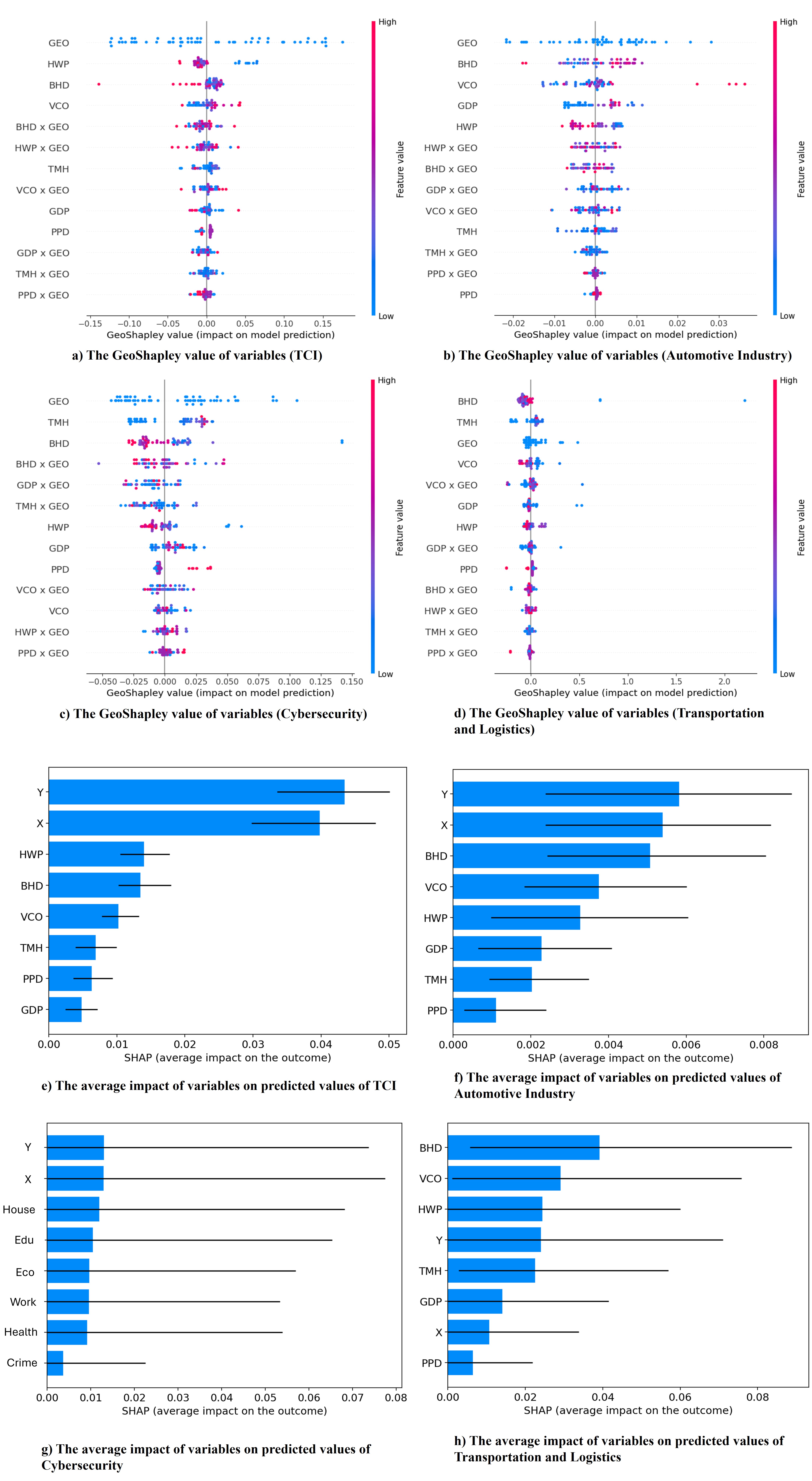}
    \caption{The average impact evaluation of social variables on the predicted visitor flows in the TCI categories}
    \label{fig:13}
\end{figure}

To further examine the influence of various factors on visitor flows within TCI, the spatial distribution of location-based contributions was analyzed using SHAP values derived from geographic coordinates. These contributions isolate the impact of location while accounting for other factors, effectively capturing contextual regional effects. Fig. \ref{fig:13} presents a summary plot of estimated SHAP values for TCI classification, with feature values represented by color. The plot ranks major contributing factors by importance, from top to bottom, revealing that geolocation exerts the strongest influence on TCI, followed by the work variable, with education and crime variables playing secondary roles. This indicates that geolocation and work-related factors are the most significant driver of visitor flows in TCI. Specifically, visitor flows can fluctuate by as much as ±0.15\% depending on location, while variations in working hours contribute to changes ranging from -0.04\% to 0.06\%.

Highly educated professionals are more likely to concentrate in economically developed regions with strong industry clusters, where higher education institutions play a key role in talent cultivation. Additionally, knowledge acquisition in these industries frequently occurs through external business interactions rather than formal educational institutions \cite{ostergaard2009knowledge}. Consequently, within TCI industries, employees with lower levels of education often engage in more cross-regional mobility to develop their skills, whereas highly educated individuals are more likely to move locally within the same economic cluster rather than relocate over long distances for job opportunities.

In the automotive industry, after geolocation, education emerges as the most influential factor (Fig. \ref{fig:13}b), while housing and health variables have minimal impact. In the cybersecurity sector, geolocation remains the most influential factor, followed by housing and education variables and their interactions with location. Visitor flows fluctuate by up to ±0.125\% depending on location, while variations in the income-to-property ratio affect visitor flows by ±0.03\%, highlighting the interplay between location-based and socioeconomic factors. In contrast, crime, work, and health variables, along with their interactions with geolocation, have minimal impact.

Compared to other industries, the impacts of all variables in the transportation and logistics sector are significantly lower, with geolocation playing a diminished role (Fig. \ref{fig:13}d). Education becomes the most influential factor, followed by housing, while other variables exert minimal influence. This suggests that educational attainment and housing affordability shape visitor flows more than geographic proximity. Although the industry shares co-location patterns with cybersecurity, geolocation-related factors contribute little, implying that broader structural elements drive its growth. Logistics industry clustering enhances regional economic growth and production efficiency, with co-location benefits varying by knowledge intensity \cite{liu2022logistics,delgado2020co}. These clusters generate economies of scale through information sharing, shared inputs, and technology spillovers, influencing spatial diffusion and sectoral distribution \cite{sun2018logistics,aljohani2016impacts}, though their indirect effects on dynamic visitor flows remain underexplored in social variable analyses.

\section{Conclusion}
In the study, we propose a novel research framework that examines how socio-economic factors in the U.S influence the distribution patterns and clustering characteristics of TCI from a unique visitor flow perspective, and we innovatively forecast future variations in visitor flow within TCI. By integrating socio-economic theory with market development phenomena, we reveal the developmental patterns of the TCI industry in the U.S in 2022. Methodologically, we combine cutting-edge DL techniques by incorporating attention mechanisms and GCN as the core components to introduce an innovative BiTransGCN model capable of accurately predicting future changes in visitor flow within TCI. Additionally, we analyze spatial clustering characteristics using the global Moran’s I index and the GeoShapley method, while assessing the contributions of key explanatory variables. Our findings reveal the intrinsic relationships between cluster levels and various critical socio-economic factors. Alongside the backdrop of evolving industrial development trends, six categories of social variables, including health, education, crime, employment, economy, and housing, significantly impact the spatial patterns of visitor flow through key mechanisms such as the co-location effect, industrial clustering, geographic economic transfer, and knowledge spillover.

We found that visitor flows and industrial clustering strongly correlate with the automotive industry's shift from Michigan and the Northeast to the Southern U.S., notably Texas, driven by global supply chains and cross-border trade with Mexico. Projections for the next industry cycle indicate differentiated growth trajectories across the three main TCI sectors. While TCI industries are expected to expand across all 51 states at an average rate of 14.16\%, growth will be driven primarily by the automotive and cybersecurity industries, whereas the transportation and logistics sector is projected to decline. This suggests an ongoing shift in the industrial structure of TCI, influenced by geographic and social factors.

Concerning potential limitations in the research. Although we normalized six categories of representative social variables to quantify regional attractiveness, the selected indicators remain incomplete; future research should incorporate additional factors such as industrial competitiveness and market size to develop a more comprehensive, multidimensional socioeconomic analysis framework. Meanwhile, the proposed BiTransGCN architecture leverages the advantages of self-attention and graph convolutional networks but introduces high computational complexity, large storage costs, and challenges in balancing global and local feature modeling, necessitating a better trade-off between efficiency and performance optimization. We look forward to improving these potential drawbacks in future research.

Overall, the proposed BiTransGCN model performs comprehensive prediction capabilities of visitor flow in TCI cluster research. TCI spatial clustering effects and spatiotemporal dynamics in the U.S. remain significant, with notable transformations occurring within industry structures. Geolocation and education are the dominant factors driving industrial clustering, and projections indicate that domestic visitor flows and clustering effects will continue to intensify, solidifying Texas as a key hub for future industry development. These findings offer valuable insights for policymakers, business leaders, and industry professionals, providing scientific guidance for strategic economic development and long-term planning within the TCI landscape in the U.S.

\bibliographystyle{unsrt}
\bibliography{Reference}

\end{document}